\newif\ifsingle
\singlefalse

\newif\ifproofs

\ifsingle
\documentclass[11pt,draftclsnofoot, onecolumn]{IEEEtran}		
\else		
\documentclass[10pt,final, twocolumn]{IEEEtran}
\fi

\usepackage{multirow}

\usepackage{soul}
\usepackage{times}
\usepackage{amsmath,dsfont}
\usepackage{amssymb,amsthm}
\usepackage{epsfig,verbatim}
\usepackage{subfigure}
\usepackage{setspace}
\usepackage{color}
\usepackage{cite}
\usepackage{epstopdf}
\usepackage{graphics}
\usepackage{accents}
\usepackage{acronym}
\usepackage[bookmarks,colorlinks]{hyperref}
\usepackage{booktabs}
\usepackage{mathtools}
\usepackage{algorithm}
\usepackage{algorithmic}
\usepackage{enumitem}

\newcommand{\myVec}[1]{{\mathbf{#1}}}
\newcommand{\myMat}[1]{{\mathbf{#1}}}
\newcommand{\mySet}[1]{\mathcal{#1}}

\newcommand{\dtmax}{N_{\rm rec}}
\newcommand{\dtidx}{n}
\newcommand{\dtrad}{N_{r}}

\newcommand{\ReviseDing}[1]{\textcolor{black}{#1}}
\newcommand{\ReviseRadar}[1]{\textcolor{black}{#1}}

\newtheorem{theorem}{Theorem}
\newtheorem{corollary}{Corollary}
\newtheorem{proposition}{Proposition}

\definecolor{NewColor}{rgb}{0,0,0} 

\ifsingle

\newcommand{\subfigWidth}{0.65\columnwidth}

\else

\newcommand{\subfigWidth}{0.7\columnwidth}

\fi 

\acrodef{adc}[ADC]{analog-to-digital convertor}
\acrodef{dac}[DAC]{digital-to-analog convertor}
\acrodef{cs}[CS]{compressed sensing}
\acrodef{dtft}[DTFT]{discrete-time Fourier transform}
\acrodef{bpsk}[BPSK]{binary phase shift keying}
\acrodef{ber}[BER]{bit error rate}
\acrodef{ofdm}[OFDM]{orthogonal frequency division multiplexing}
\acrodef{csi}[CSI]{channel state information}
\acrodef{map}[MAP]{maximum a-posteriori probability}
\acrodef{snr}[SNR]{signal-to-noise ratio}
\acrodef{bs}[BS]{base station} 
\acrodef{mimo}[MIMO]{Multiple-input multiple-output}
\acrodef{mse}[MSE]{mean-squared error}
\acrodef{pdf}[PDF]{probability density function}
\acrodef{rv}[RV]{random variable}
\acrodef{lti}[LTI]{linear time-invariant}
\acrodef{wss}[WSS]{wide-sense stationary}
\acrodef{psd}[PSD]{power spectral density}
\acrodef{ser}[SER]{symbol error rate} 
\acrodef{isi}[ISI]{intersymbol interference} 
\acrodef{lstm}[LSTM]{long short-term memory} 
\acrodef{em}[EM]{expectation minimization} 
\acrodef{reg}[REG]{radar echo generator} 
\acrodef{tdd}[TDD]{time division duplexing} 
\acrodef{ut}[UT]{user terminal} 
\acrodef{awgn}[AWGN]{additive white Gaussian noise}
\acrodef{cgac}[CGAC]{Complex-gain analog combiner}
\acrodef{psoac}[PSOAC]{Phase-shifter-only analog combiner}
\acrodef{fpga}[FPGA]{field-programmable gate array}
\acrodef{gui}[GUI]{graphical user interface}
\acrodef{dfrc}[DFRC]{dual function radar-communications}
\acrodef{jrc}[JRC]{joint radar and communications}
\acrodef{pri}[PRI]{pulse repetition interval}
\acrodef{gsm}[GSM]{generalized spatial modulation}
\acrodef{smx}[SMX]{spatial multiplexing MIMO}
\acrodef{lfm}[LFM]{linear frequency modulation}
\acrodef{tws}[TWS]{track while scan}
\acrodef{stt}[STT]{single target tracker}
\acrodef{mi}[MI]{mutual information}
\acrodef{ula}[ULA]{uniform linear array}
\acrodef{scr}[SCR]{signal-to-clutter ratio}
\acrodef{im}[IM]{index modulation}
\acrodef{spacora}[SpaCoR]{spatial modulation based communication-radar}
\acrodef{majorcom}[MAJoRCom]{multi-carrier agile joint radar communication}
\acrodef{omp}[OMP]{orthogonal matching pursuit}

\IEEEoverridecommandlockouts

\title{Spatial Modulation for Joint Radar-Communications Systems: Design, Analysis, and Hardware Prototype
}

\author{
	\IEEEauthorblockN{Dingyou Ma, Nir Shlezinger,  Tianyao Huang, Yariv Shavit, Moshe Namer, Yimin Liu,  and Yonina C. Eldar
	} 
	\thanks{Parts of this work were presented in the 2018 IEEE International Conference on Acoustics, Speech, and Signal Processing (ICASSP) as the paper \cite{Ma2018A}.
	}
	\thanks{
		D. Ma, T. Huang, and Y. Liu are with the EE Department, Tsinghua University, Beijing, China (e-mail: mdy16@mails.tsinghua.edu.cn; \{huangtianyao, yiminliu\}@tsinghua.edu.cn).
		N. Shlezinger, Y. Shavit,  and Y. C. Eldar are with the Faculty of Math and CS, Weizmann Institute, Rehovot, Israel (e-mail: \{nir.shlezinger, yariv.shavit, yonina.eldar\}@weizmann.ac.il). 	
		M. Namer is with the EE Department, Technion, Haifa, Israel (e-mail:namer@ee.technion.ac.il). 	
		This work received funding from the National Natural Science Foundation of China under grant 61801258, from the European Union’s Horizon 2020 research and innovation program under grant No. 646804-ERC-COG-BNYQ, and from the Air Force Office of Scientific Research under grant No. FA9550-18-1-0208. (\emph{Corresponding author: Tianyao Huang})
	}		
	\vspace{-0.5cm}
}
\vspace{-0.75cm}

\begin{document}
	
	\maketitle
	\pagestyle{plain}
	\thispagestyle{plain}
	\begin{abstract} 
		Dual-function radar-communications (DFRC) systems implement  radar and communication functionalities on a single platform. Jointly designing these subsystems can lead to substantial gains in performance as well as size, cost, and power consumption. In this paper, we propose a DFRC system, which utilizes generalized spatial modulation (GSM) to realize coexisting radar and communications waveforms. Our proposed GSM-based scheme, \ReviseDing{referred to as spatial modulation based communication-radar (SpaCoR) system}, allocates antenna elements among the subsystems based on the transmitted message, thus achieving increased communication rates by embedding additional data bits in the antenna selection. We formulate the resulting signal models, and present a dedicated radar processing scheme. To evaluate the radar performance, we characterize the statistical properties of the transmit beam pattern. Then, we present a hardware prototype of the proposed DFRC system, demonstrating the  feasibility of the scheme. Our results show that the proposed GSM system achieves improved communication performance compared to techniques  utilizing fixed allocations operating at the same data rate. For the radar subsystem, our experiments show that the spatial agility induced by the GSM transmission  improves the angular resolution and reduces the sidelobe level in the transmit beam pattern \ReviseDing{compared to using fixed antenna allocations.}
	\end{abstract}
	
	
	\vspace{-0.4cm}
	\section{Introduction} 
	\vspace{-0.1cm}
	A wide variety of systems, ranging from autonomous vehicles to military applications, implement both radar and communications. Traditionally, these two functionalities are designed independently, using separate subsystems.  An alternative strategy, which is the focus of growing research attention, is to {\em jointly design} them as a \ac{dfrc} system  \cite{Ma2020Joint, Paul2017, Tavik2005Advanced, han2013joint, zheng2019radar}. Such joint designs improve performance by facilitating coexistence \cite{Paul2017}, as well as contribute to reducing the number of antennas \cite{Tavik2005Advanced},  system size, weight, and power consumption \cite{Liu2017a}. 
	
	A common approach for realizing \ac{dfrc} systems utilizes  a single dual-function waveform, which is commonly based on traditional  communications signaling,  or on an optimized joint waveform \cite{Sturm2009An,Liu2017Towards, Liu2018}.
	The application of 
	\ac{ofdm} communication signaling for probing  was studied in  \cite{Sturm2009An}, while employing spread spectrum waveforms for \ac{dfrc} systems was considered in \cite{Sturm2011Waveform}. 
	The usage of such signals, which were originally designed for communications, as dual function waveforms, inherently results in some performance degradation \cite{Chiriyath2017}. For instance, using \ac{ofdm} signaling leads to waveforms with high peak-to-average-power ratio, which induces distortion in the presence of practical power amplifiers, and limits the radar detection capability in short ranges \cite{ZhangYu2017Waveform}. 
	\ac{mimo} radar, which transmits multiple waveforms simultaneously, facilitates designing optimized dual-function waveforms\cite{Liu2017Towards, Liu2018}. These optimized waveforms balance the tradeoff between  communication and radar performance in light of the  constraints imposed by both systems. However, such joint optimizations require prior knowledge of the communication channel and the radar targets, which is likely to be difficult to acquire in dynamic setups, and typically involves solving a computationally complex optimization problem. 
	
	
	When radar is the primary user, a promising \ac{dfrc} method is to embed the message into the radar waveform via \ac{im} \cite{Basar2016Index}. In \ac{mimo} radar, \ac{im} can be realized by conveying the information  in the radar  sidelobes \cite{Hassanien2016b}, using frequency hopping waveforms \cite{Hassanien2017}, and in the permutation of orthogonal waveforms among the  elements \cite{Wang2018Dual}. Recently, the work \cite{Huang2020MAJoRCom} proposed the \ac{im}-based   \ReviseDing{\ac{majorcom} system}, which \textcolor{black}{embeds the communication message into the transmission parameters of  frequency and spatially agile radar waveforms}. While these techniques  induce minimal effect on  radar performance, they typically result in low data rates compared to using dedicated communication signals.
	
	\ReviseDing{\ac{dfrc} strategies utilizing a single waveform inherently induce performance loss on either its radar functionality, as in \ac{ofdm} waveform based methods, or lead to a low communication rate,  which is the case with the radar waveform based \ac{majorcom}. An alternative \ac{dfrc} strategy  is to utilize independent radar and communication waveforms, allowing each functionality to utilize its suitable signaling method.}
	When using individual waveforms, one should facilitate coexistence by controlling their level of mutual interference. This can be achieved by using fixed non-overlapping bands and antennas \cite{Tavik2005Advanced}, as well as by efficiently allocating bandwidth resources between the subsystems  \cite{bicua2019multicarrier}.
	In \ac{mimo} radar, coexistence can be achieved  by beamforming each signal in the proper direction \cite{McCormick2017Simul,liu2020joint} as well as by using spectrally and spatially orthogonal waveforms \cite{Liu2018}. 
	The resulting tradeoff between radar and communication of this strategy stems from their mutual interference as well as the resource sharing between the subsystems, in terms of spectrum, power, and antennas. 
	


	\textcolor{black}{In this work we propose a  \ac{spacora} system, which implements a mixture of individual radar and communications waveforms with \ac{im} via \ac{gsm} \cite{Younis2010Generalised,Wang2012Generalised}. }  \ac{gsm} combines \ac{im}, in which data is conveyed in the transmission parameters, with dedicated communications signaling. As such, the proposed approach exhibits only a minor degradation in radar performance due to the presence of data transmission, as common in \ac{im} based \ac{dfrc} systems \cite{Ma2020Joint}, 
	while supporting the increased data rates with individual waveforms. \textcolor{black}{We demonstrate the feasibility of \ac{spacora} by presenting a hardware prototype which implements this scheme.}
	
	In particular, we consider a system in which radar and communications use different fixed bands, thus complying with existing  standardization. To avoid the hardware complications associated with transmitting multiband signals, we restrict each antenna element to transmit only a single waveform, either radar or communications. To maximize the performance under these restrictions, the proposed method  allocates the antenna array elements between the radar and communications subsystems, which operate at different bands thus avoiding mutual interference. The allocation is based on the transmitted message using \ac{gsm}, thus embedding some of the data bits in the antenna selection, inducing spatial agility \cite{Huang2019b}. As the communications subsystem is based on conventional \ac{gsm}, for which the performance was theoretically characterized in \cite{Younis2018Information}, we analyze only the radar performance of  \ReviseDing{\ac{spacora}}. In particular, we prove that  its agile profile mitigates the degradation in radar beam pattern due to using a subset of the antenna array, which in turn improves its accuracy over approaches with a fixed antenna allocation. 
	
	We implement  \ReviseDing{\ac{spacora}} in a specifically designed hardware prototype utilizing a two-dimensional antenna with 16 elements, demonstrating the practical feasibility of the proposed \ac{dfrc} system. Our prototype allows to evaluate  \ReviseDing{\ac{spacora}}  using actual passband waveforms with over-the-air signaling. In our experimental study, we compare  \ReviseDing{\ac{spacora}}  to \ac{dfrc} schemes using individual subsystems with fixed antenna allocation. Our results show that the communications subsystem of \ReviseDing{\ac{spacora}} achieves improved \ac{ber}  performance compared to the fixed allocation system when using the same data rate. For the radar subsystem, our experiments show that spatial agility of  \ReviseDing{\ac{spacora}}  leads to improved angular resolution.

	The rest of the paper is organized as follows: Section~\ref{sec:Model} presents the system model, detailing the communications and radar subsystems. Section \ref{sec:RadarPerformance}  analyzes the  radar transmit  beam pattern. 
	The high level design  of the \ac{dfrc} system prototype is described in Section \ref{sec:Highleveldesign}, and the  implementation of each of its components  is detailed in Section \ref{sec:ProtoRealization}. We evaluate the performance of the proposed system in a set of experiments in Section~\ref{sec:Sims}. Finally, Section \ref{sec:Conclusions} provides concluding remarks.
	
	The following notations are used throughout the paper: Boldface lowercase and uppercase letters denote vectors and matrices, respectively.  
	We denote the transpose, complex conjugate, Hermitian transpose and integer floor operation as $\left(\cdot\right)^{\mathrm{T}}$,   $\left(\cdot\right)^{\ast}$,   $\left(\cdot\right)^{\mathrm{H}}$ and $\lfloor \cdot \rfloor$,  respectively. The complex normal distribution with mean $\mu$ and variance $\sigma^2$ is expressed as $\mathcal{CN}\left(\mu, {\sigma}^2\right)$, while $\mathcal{E}\left\{\cdot\right\}$ and $\mathcal{V}\left\{\cdot\right\}$ are the expected value and variance  of a random argument, respectively.   The sets of complex and natural numbers are $\mathbb{C}$ and  $\mathbb{N}$, respectively. 
	\vspace{-0.2cm}
	\section{SpaCoR System  Model}
	\label{sec:Model}
	\vspace{-0.1cm}
	Here, we detail the proposed \ReviseDing{\ac{spacora}} system. To that aim, we first discuss the main guidelines and model constraints under which the system is designed in Subsection~\ref{subsec:Constraints}. Then, in Subsection~\ref{subsec:DFRCModel} we present the overall \ac{dfrc} method, and elaborate on the individual communications and radar subsystems in Subsections \ref{subsec:Comm}-\ref{subsec:Radar}, respectively.

	\vspace{-0.2cm}
	\subsection{System Design Guidelines and Constraints}
	\label{subsec:Constraints}
\begin{figure}
		\centerline{\includegraphics[width=\columnwidth]{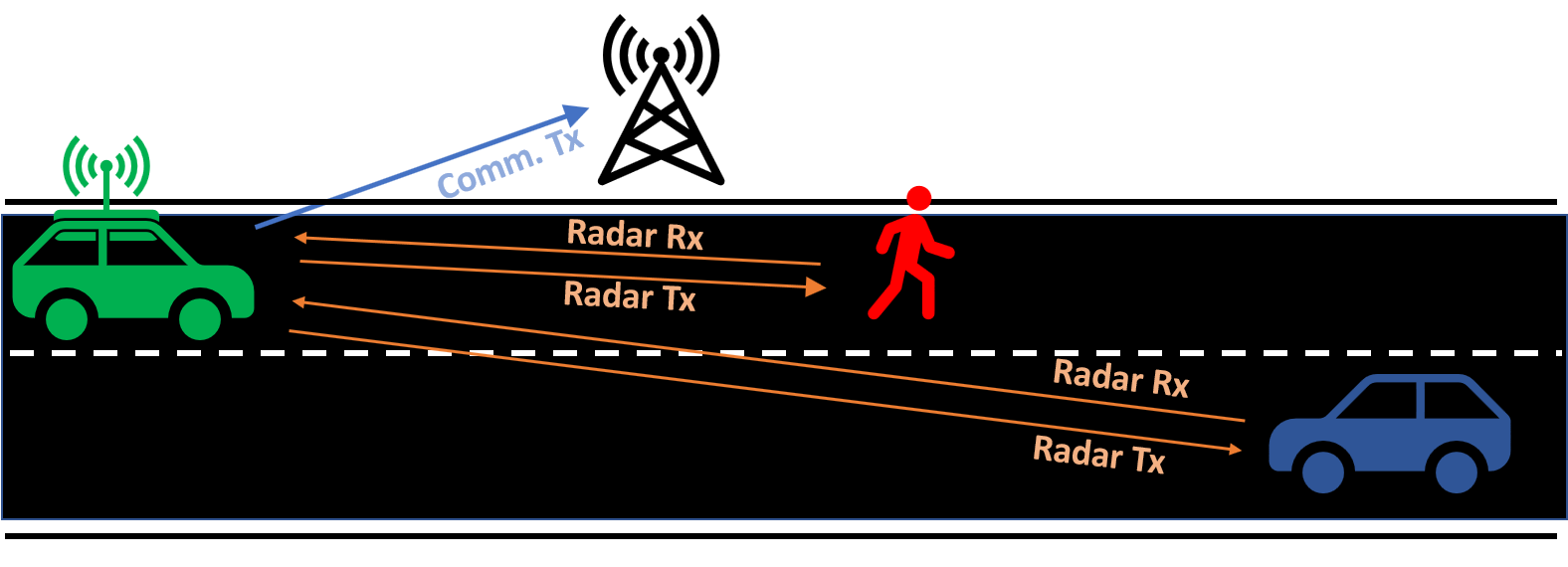}}
		\vspace{-0.2cm}
		\caption{An autonomous vehicle with radar and communications functionalities.}
		\label{fig:DFRC1}
	\end{figure}

	\vspace{-0.1cm}
	We consider a system equipped with a phased array antenna implementing active radar sensing while communicating with a remote receiver. An illustration of such a system in the context of vehicular applications is given in Fig. \ref{fig:DFRC1}. 
	In the \ac{dfrc} system, radar is the primary user and communications is the secondary user.  We consider a pulse radar, in which the transmission and reception are carried out in a  time-division duplex manner. The communication signal is transmitted during radar transmission. 
	Only the communications receiver is required to have \ac{csi}, while the \ac{dfrc} system can be ignorant of the channel realization.  
	
	We require the radar and communication functionalities to operate on the same antenna array without mutual interference. An intuitive approach to implement such orthogonality is by time sharing. However, for many applications, radar needs to work continuously in time, rendering time sharing 
	irrelevant. An alternative approach is to boost spatial orthogonality by beamforming, as in, e.g., \cite{McCormick2017Simul, liu2020joint}. However, these approaches typically require fully configurable \ac{mimo} arrays as well as knowledge of the communication channel. Consequently, we set the subsystems to use non-overlapping frequency bands, allowing these functionalities to work simultaneously in an orthogonal fashion while complying with conventional communication standards and spectral allocations.

	Finally, in order to maintain high power efficiency, we avoid the transmission of multiband signals. Consequently, each antenna element can only be utilized for either radar or communications signalling at a given \ac{pri}. By doing so, each element transmits narrowband signals, avoiding the envelop fluctuations and reduction in power efficiency associated with multiband signaling \cite{Huang2019b}. 
	
	To summarize, our system is designed to comply with the following guidelines and constraints:
	\begin{itemize}
		\item Radar is based on pulse probing. 
		\item The same array element cannot be simultaneously used for both radar and communications transmission.
		\item Both functionalities transmit at the \ReviseDing{same} time,
		and the returning radar echoes are captured in the complete array. 
		\item The waveforms are orthogonal in spectrum.
		\item The communications subsystem  operates without \ac{csi}.
	\end{itemize}
	
	An intuitive design approach in light of the above constraints is to divide the antenna array into two fixed sub-arrays, each assigned to a different subsystem, resulting in separate systems. Nonetheless, we next show that performance gains in both radar and communications  can be achieved using a joint design,  which guarantees a low complexity structure, while complying with the aforementioned constraints.  
	
	\vspace{-0.2cm}
	\subsection{SpaCoR System}
	\label{subsec:DFRCModel}
	\vspace{-0.1cm}
	To formulate the proposed \ac{dfrc} method, we first elaborate on the drawbacks of using fixed allocation, after which we discuss how these drawbacks are tackled in our joint design. We focus on systems equipped with a \ac{ula}  consisting of $M$ antenna elements with inter-element spacing $d$.
	%
	\textcolor{black}{The antenna array is a one-dimensional element-level digital array, where the transmit waveforms are generated digitally for each element,  
		facilitating beamforming in digital baseband. While here we focus on one-dimensional arrays for ease of presentation, our prototype detailed in Section \ref{sec:Highleveldesign} uses a two-dimensional antenna surface.} 
	
\begin{figure*}
		\centerline{\includegraphics[width=1.7\columnwidth]{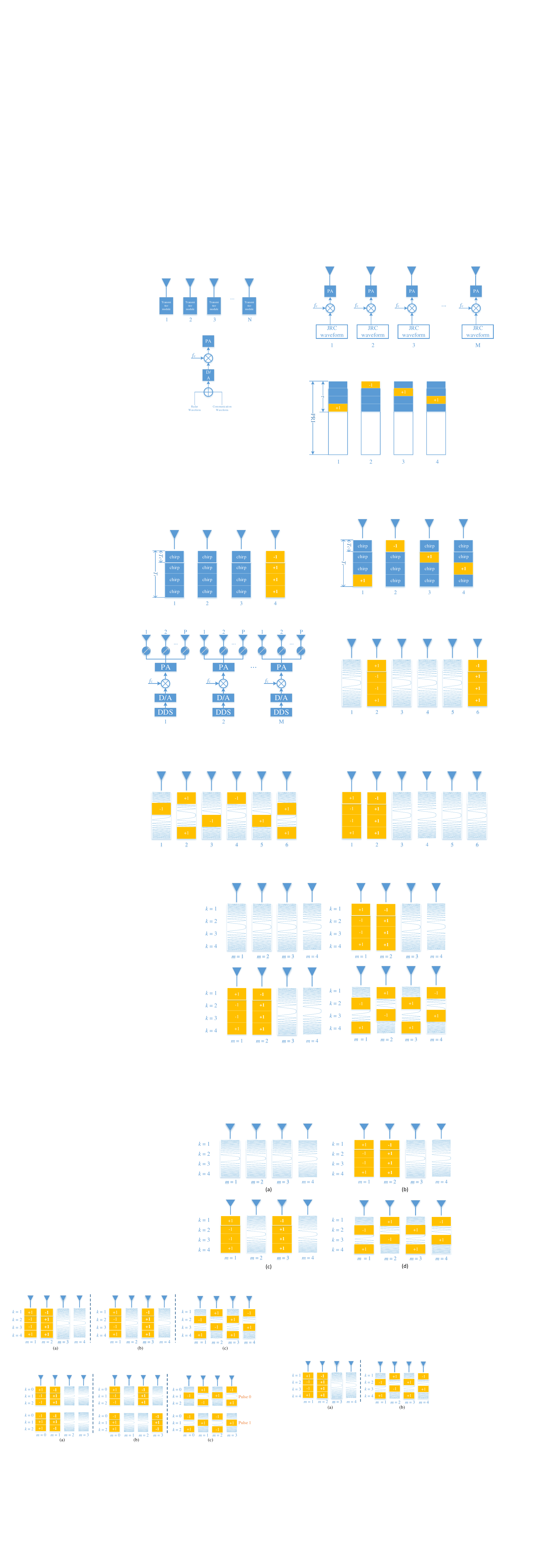}}
		\vspace{-0.3cm}
		\caption{Illustration of the fixed antenna allocation schemes and \ReviseDing{\ac{spacora}}. The chirplet waveform represents radar transmission while $\pm 1$ denotes a communication symbol. (a) \emph{Fix1}: The antenna array is divided into two sub-\ac{ula}s, and the allocation pattern remains static.  (b) \emph{Fix2}: 
			The  allocation pattern changes randomly between different radar pulses. (c) \ReviseDing{\ac{spacora} }: Here, the allocation varies  between symbol time slots. 
		}
		\vspace{-0.3cm}
		\label{fig:AntennaAlocationSchemes}
	\end{figure*}

	As mentioned in the previous subsection, an intuitive approach is to divide the antenna elements between radar and communications in a fixed manner such that the antenna allocation pattern is static during each radar pulse duration. One simple fixed allocation scheme is obtained by dividing the antenna array into two sub-\ac{ula}s, referred to henceforth as \emph{Fix1} and illustrated in Fig. \ref{fig:AntennaAlocationSchemes}(a). Another fixed allocation approach randomly divides the antenna array into two sub-arrays while allowing the allocation pattern to change between different radar pulses. This technique is referred to as \emph{Fix2}, and is illustrated in Fig. \ref{fig:AntennaAlocationSchemes}(b). In these fixed antenna allocation methods, during each pulse transmission, $K$ symbols are transmitted from the antenna elements assigned to communications. 
	In the example in Fig. \ref{fig:AntennaAlocationSchemes}, two antenna elements are allocated for communication in a static manner during each radar pulse, and each element transmits $K=3$ symbols during one radar pulse, while the remaining two elements are allocated to radar. 
	
	These fixed allocation techniques affect the performance of both radar and communications. Compared with traditional phased array radar utilizing all the elements for radar transmission, using a fixed sub-array yields a wider mainlobe or higher sidelobes in the transmit beam pattern, which are shown in Section  \ref{sec:RadarPerformance}.
	For the communications subsystem, fixed allocation does not exploit the fact that the system is equipped with a larger number of elements than what is actually utilized, and the data rate can be increased by exploiting the spatial diversity.
	
	
	To exploit the full antenna array for both radar and communications, we propose a \ac{dfrc} scheme which randomly allocates the antenna  elements between radar and communications. During  transmission, the antenna allocation is changed  between different symbol slots. Inspired by \ac{gsm} communications \cite{Younis2010Generalised,Wang2012Generalised}, the selection of the specific antennas is determined by some of the bits intended for transmission. 
	
	\ReviseDing{\ac{spacora}} overcomes both the radar and communications drawbacks of using fixed allocation schemes: For the radar subsystem, each element is effectively used for probing with high probability over a large number of time slots, which results in a transmit beam pattern approximating the beam pattern of the full antenna array.
	In fact, as we show in our analysis in Section \ref{sec:RadarPerformance}, the resulting expected beam pattern approaches that achieved when using the full array for radar. For the communication functionality,  additional bits are conveyed in the selection of the antennas. These additional bits increase the data rate, or alternatively, allow the usage of \ReviseDing{sparser}  constellations in the dedicated waveform compared to fixed allocation with the same data rate.    
	An illustration of the resulting waveform is depicted in Fig.~\ref{fig:AntennaAlocationSchemes}(c). In this example, two antennas are allocated for communication at every time instance. 
	The information bits are conveyed by the combination of the communications antennas and via the  signals transmitted from them, as detailed in the sequel.
	
	\vspace{-0.2cm}
	\subsection{Communications Subsystem}
	\label{subsec:Comm}
	\vspace{-0.1cm}
	The proposed communications subsystem, which utilizes dedicated waveforms while allowing extra information bits to be conveyed in the selection of transmit antennas, implements \ac{gsm} signaling  \cite{Younis2010Generalised,Wang2012Generalised}. Therefore, to formulate the communications subsystem, we start with a brief review of \ac{gsm}, after which we discuss how the received symbols are decoded. 
	
	\subsubsection{Generalized Spatial Modulation}
	\ac{gsm}, originally proposed in \cite{Younis2010Generalised}, combines spatial \ac{im} with multi-antenna transmission, aiming at increasing the data rate when using a subset of the antenna array elements.
	\ReviseRadar{As such, \ac{gsm} refers to a family of \ac{im}-based methods. Our proposed \ac{dfrc} system specifically builds upon the \ac{gsm} scheme of  \cite{Wang2012Generalised}, as detailed next.}
	
	The information bits conveyed in each \ac{gsm} symbol are divided into spatial selection bits and constellation bits. The spatial selection bits determine the indices of the transmit antennas. 
	By letting $M^{c}_{T} < M$ be the number of antennas used for communications transmission, it holds that there are $\binom{M}{M^{c}_{T}}$ different possible antenna combinations. As a result, $\lfloor \log_2 \binom{M}{M^{c}_{T}} \rfloor$ bits can be conveyed through the antenna selection in each \ac{gsm} symbol. The selected antennas are used to transmit the symbols embedding the constellation bits. While, in general, \ac{gsm} can be combined with any form of signaling \cite{Wang2012Generalised}, we focus on phase shift keying to maintain constant modulus waveforms. When a constellation $\mySet{J}$ of cardinality $|\mySet{J}| = J$ is utilized, $R = M^{c}_{T}\log_2 J + \lfloor \log_2\binom{M}{M^{c}_{T}} \rfloor$ uncoded bits are conveyed in each \ac{gsm} symbol. Compared with fixed antenna allocation approaches with the same constellation order, \ac{gsm} enables $\lfloor \log_2\binom{M}{M^{c}_{T}} \rfloor$ additional bits to be embedded in each symbol. The transmission does not require knowledge of the underlying communication channel. When such \ac{csi} is available, it can be exploited by, e.g., spatial precoding \cite{perez2010mimo}.
	
\begin{figure} 
		\centerline{\includegraphics[width=0.75\columnwidth]{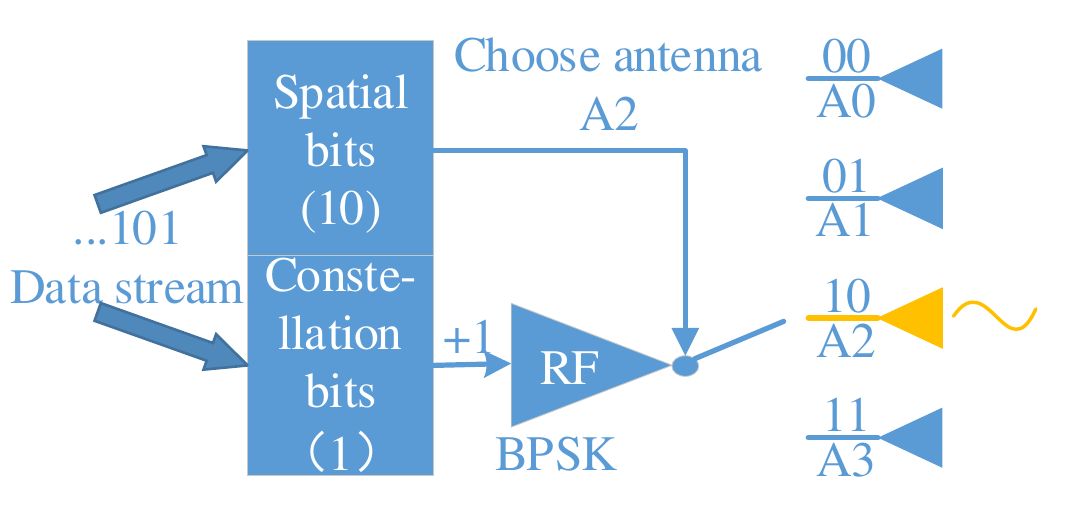}}
		\vspace{-0.3cm}
		\caption{\ac{gsm} transmission illustration.}
		\label{fig:GSMmodel}
	\end{figure}

	An example of \ac{gsm} transmission is shown in Fig. \ref{fig:GSMmodel}. In this example, the antenna array has $M = 4$ elements. A single antenna is used for each symbol, i.e., $M^{c}_{T} = 1$ and $\lfloor \log_2\binom{4}{1} \rfloor = 2$ bits are embedded in the combination of transmit antennas. A \ac{bpsk} modulation is utilized, thus $\mySet{J} = \{\pm 1\}$, $J = 2$, and a total of $R = 3$ bits are conveyed in each symbol. In the example in
	Fig. \ref{fig:GSMmodel}, the  message $101$  is divided into spatial selection bits $10$ and constellation bit $1$. According to the element mapping rule, antenna A2 transmits the \ac{bpsk} symbol $+1$.
	
	In  \ac{gsm} signalling, only a subset of the antenna array is used, and the transmit elements  change between different symbols. Hence, the remaining elements can be assigned to radar transmission, leading to the proposed \ac{gsm}-based \ac{dfrc} system, which complies with the constraints discussed in Subsection \ref{subsec:Constraints}. As each radar pulse consists of $K$ symbol slots, 
	a total of $K \cdot R$ bits are conveyed in each \ac{pri}. 
	
	\subsubsection{Communications Receiver Operation} 
	To formulate how the transmitted signal is decoded by the receiver, we first model the channel output. In the following we consider a \ac{mimo} receiver with $M_{R}^{c}$ antennas, and assume that, unlike the \ac{dfrc} transmitter, it has full \ac{csi}.
	
	Since radar and communications use distinct bands,  no cross interference exists. Consequently, by letting  $\mathbf{x} \in \mySet{X} \subset (\mySet{J}\cup \{0\}) ^{M}$ denote the channel input at the communication frequency range, it holds that $\mathbf{x}$ is sparse with support size $M_{T}^{c}$, i.e., $ \mySet{X}$ is the set of $M^{c}_{T}$ sparse vectors in  $(\mySet{J}\cup \{0\}) ^{M}$. Therefore, assuming a linear memoryless channel  $\mathbf{H} \in \mathbb{C}^{M_{R}^{c}\times M}$ whose output is corrupted by an additive noise vector $\mathbf{w}^{\left(c\right)} \in \mathbb{C}^{M_{R}^{c} }$, the channel output representing a single \ac{gsm} symbol observed by the receiver, denoted as $\mathbf{y}^{\left(c\right)} \in \mathbb{C}^{M_{R}^{c}}$, is given by   
	$\mathbf{y}^{\left(c\right)} = \mathbf{Hx} + \mathbf{w}^{\left(c\right)}$.
	Since the receiver has full \ac{csi}, i.e., knowledge of the matrix  $\mathbf{H}$ and the distribution of $\mathbf{w}^{\left(c\right)}$, it can decode with minimal probability of error using the maximum a-posteriori probability rule. Assuming that the data bits are equiprobable, this symbol detection rule is given by 
	\begin{equation}
	\label{eqn:ML}
	\hat{\mathbf{x}} = \mathop {\arg \max}\limits_{\mathbf{x} \in \mySet{X}} p\left(\mathbf{y}^{\left(c\right)}|\mathbf{x},\mathbf{H}\right).
	\end{equation}
	For example, when the noise obeys a  white Gaussian distribution,  
	\eqref{eqn:ML} specializes to the minimum distance detector. 
	
	Given the detected $\hat{\mathbf{x}}$, the spatial selection bits can then be recovered from the support of $\hat{\mathbf{x}}$, while the constellation bits are demodulated from its non-zero entries.
	Recovering $\hat{\mathbf{x}}$ via \eqref{eqn:ML} generally involves searching over the set $\mySet{X}$ whose cardinality is $R$. \textcolor{black}{When $R$ is large, symbol detection can be  facilitated using reduced complexity  \ac{gsm} decoding methods, see e.g., \cite{Wang2012Generalised}, allowing \ac{spacora} to be utilized with controllable decoding complexity  at the receiver side. In our experimental   study detailed in Section \ref{sec:Sims} we consider scenarios in which $R$ is relatively small, and thus carry out symbol detection by directly computing \eqref{eqn:ML}.} 
	
	
	\vspace{-0.2cm}
	\subsection{Radar Subsystem}
	\label{subsec:Radar}
	\vspace{-0.1cm}	
	To formulate the radar subsystem, in the following we first model its transmitted and received signal, after which we introduce an algorithm for radar detection.

	\ac{spacora}  uses a phased array radar, which enables to steer the radar beam at the direction of interest. Such beam steering is achieved by using a single waveform, denoted by $s(t)$, while assigning a different weight per each element designed to steer the beam in a desired direction $\theta_T$. \ReviseRadar{The transmit power  is denoted as $P_t:= \frac{1}{T_{\rm PRI}} \int_{0}^{T_{\rm PRI}} \left|s\left(t\right)\right|^2 dt$, where $T_{\rm PRI}$ is the pulse repetition interval.} 
	For a \ac{ula} with $M$ elements, 
	the weight function 
	of the $m$th element is $a_{m}\left(\theta_{T}\right) = e^{-j\frac{2\pi m d\sin\theta_{T}}{\lambda}}$, where $\lambda$ is the wavelength, and the corresponding waveform 
	is $s_{m}\left(t\right) = s\left(t\right)a_{m}\left(\theta_{T}\right)$. For a narrowband waveform,
	the signal received in the far field at angle $\theta$ and range $\xi$ is expressed as \cite[Ch. 8.2]{Skolnik2001Inroduction}
	\vspace{-0.1cm}
	\begin{equation}
	y_{\theta}^{\left(r\right)}\left(t\right) = s\left(t - \frac{\xi}{c}\right)\sum_{m=0}^{M-1}a_{m}\left( \theta_{T}\right)a^{\ast}_{m}\left(\theta\right),
	\vspace{-0.1cm}	
	\label{eqn:PArray}
	\end{equation}
	where $a^{\ast}_{m}\left(\theta\right)$
	is the steering weight
	between the $m$th element and the far field target at angle $\theta$, and $c$ is the speed of light. 
	
	Unlike traditional phased array radar, which utilizes the complete antenna array,  { \ac{spacora}} assigns only a subset of the antenna elements for radar signalling at each time instance, and the antenna allocation pattern dynamically changes between different communication symbols. 
	In particular, letting $T_c$ be the duration of a communication symbol, each radar pulse consists of $K$ consecutive communication symbols, i.e., the pulse width is $T_r = K T_c$. At each time slot, $M_T^{r} = M - M_T^{c}$ antenna elements are assigned for radar transmission. Thus, by letting $m_{k,0}<m_{k,1}<\cdots<m_{k,M_T^{r} - 1}$ be random variables representing the element indices assigned to radar at the $k$th time slot, the received signal, which for conventional phased array is given by \eqref{eqn:PArray}, is expressed as
	\begin{equation}
	\!\!\!y_{\theta}^{\left(r\right)} \! \left(t\right)\! = \! s \! \left( \!t \! - \! \frac{\xi}{c} \!\right) \! \sum_{k=0}^{K-1} \!  g \! \left( \! \frac{t \! - \! kT_c  \! - \! \frac{\xi}{c}}{T_c} \! \right) \tilde{\rho}_{T}\left(k,\theta\right), \!\!\!
	\label{eqn:PArrayGSM}
	\end{equation}
	where $g\left(t\right)$ is a rectangular window of unity support,
	\color{black}
	and $\tilde{\rho}_T\left(k, {\theta}\right) := \sum_{l=0}^{M_T^r - 1}a_{m_{k,l}}\left(\theta_T\right)\cdot a^{\ast}_{m_{k,l}}\left( \theta\right)$ 	is the transmit gain at the $k$th time slot.  
	Denote ${\vartheta}:= \frac{2\pi d \sin\theta}{\lambda}$ and ${\vartheta_{T}}:= \frac{2 \pi d \sin\theta_{T}}{\lambda}$ as the spatial frequencies at the direction of radar target and the direction of the transmit beam, respectively. The transmit beam pattern $\tilde{\rho}_{T}\left(k, \theta\right)$ can be rewritten as
	\vspace{-0.1cm}
	\begin{equation}
	\tilde{\rho}_T\left(k, {\vartheta}\right)
	=\sum_{l=0}^{M_T^r - 1}e^{j m_{k,l}\left(\vartheta - {\vartheta_{T}}\right)}.
	\label{eqn:RhoT}
	\vspace{-0.2cm}
	\end{equation}
	\color{black}
	

	The time delay experienced by the echoes reflected from the radar target until reaching the $m$th receive antenna element is $\frac{\xi}{c} - \frac{m \vartheta}{2 \pi}$. After frequency down conversion by mixing with the local carrier, the echo received in the $m$th antenna element can be expressed as 
	\vspace{-0.1cm}
	\begin{equation}
	\!\!\!y_{m}^{\left(r\right)}\!\left(t\right) \! = \! \alpha y_{\theta}^{\left(r\right)}\! \left( \! t \! - \! \frac{\xi}{c} \! + \! \frac{m \vartheta}{2 \pi} \! \right)\!  e^{-j2{\pi}f_c t}\! +\!w_{m}^{\left(r\right)}\!\left(t\right),\!
	\vspace{-0.1cm}
	\label{eqn:RxEcho1}
	\end{equation}
	where $\alpha$ is the reflective factor of the target, $w_{m}^{\left(r\right)}\left(t\right)$ is the  noise at the $m$th antenna receiver, modeled as a white Gaussian process. 	Let $h\left(t\right)$ be the baseband radar waveform and $f_c$ denote the carrier frequency, i.e., $s\left(t\right) = h(t)e^{j2{\pi} f_c t}$. \textcolor{black}{With the  narrowband assumption, i.e., $\frac{Md}{c}\ll \frac{1}{B_{r}}$, where $B_{r}$ is the bandwidth of the radar waveform, one can use the approximation $y_{\theta}^{\left(r\right)}\! \left( \! t \! - \! \frac{\xi}{c} \! + \! \frac{m \vartheta}{2 \pi} \! \right)\approx y_{\theta}^{\left(r\right)}\! \left( \! t \! - \! \frac{\xi}{c} \right)e^{jm\vartheta}$.  Substituting \eqref{eqn:PArrayGSM} into \eqref{eqn:RxEcho1}, and defining the round trip delay  $\tau := 2\xi/c$, it follows that}
	\vspace{-0.1cm}
	\begin{equation} 
	y_{m}^{\left(r\right)}\left(t\right)=  \alpha h_m(t, \tau, \vartheta)+ w_{m}^{\left(r\right)}\left(t\right),
	\vspace{-0.1cm}
	\label{eqn:RxSignalAntm}
	\end{equation}
	where 
	\begin{align}
	&{h}_{m}\left(t, {{\tau}},  {\vartheta}\right) := e^{-j2\pi f_c {{\tau}} + j m  {\vartheta}}\notag \\
	& \qquad \times \sum_{k= 0}^{K-1}\tilde{\rho}_{T}\left(k,  {\vartheta}\right)g\left(\frac{t- kT_c - {{\tau}}}{T_c}\right)h\left(t-{{\tau}}\right).
	\label{eqn:MatchedFilter}
	\end{align}

	The radar ehco is received at the idle time of the pulse, the duration of which is $T_{\rm rec} := T_{\rm PRI} - T_{r}$. \textcolor{black}{The received signal is uniformly sampled with Nyquist rate $F_{s}$, i.e., $F_{s} \ge B_{r}$. While it is possible to sample below the Nyquist rate by using generalized sampling methods as was done in \cite{Cohen2018summer, Cohen2018SubNyquist}, we leave the study of \ac{spacora} with sub-Nyquist sampling for future work.}
	%
	The number of sample points in each \ac{pri} is $\dtmax = \lfloor \frac{T_{ \rm rec}}{T_{s}} \rfloor$, and the sample time instances are $t = \dtidx T_{s}$, where $\dtidx \in \left\{ 0,1, \cdots, \dtmax-1 \right\}$. 
	By defining \ReviseDing{$h_m\left[\dtidx , \tau, \vartheta\right]:=  h_m\left(\dtidx T_{s}, \tau, \vartheta\right)$}, 
	and $w_{m}^{\left(r\right)}\left[\dtidx\right] := w_{m}^{\left(r\right)}\left( \dtidx T_{s}\right)$, the sampled signal vector \eqref{eqn:RxSignalAntm} is given by
	\begin{equation}
	y_{m}^{\left(r\right)}\left[\dtidx\right] :=  \alpha h_m\left[\dtidx , \tau, \vartheta\right]+ n_{m}^{\left(r\right)}\left[\dtidx\right].
	\label{eqn:RxSignalAntmDiscrete}
	\end{equation}

	The discrete-time model in \eqref{eqn:RxSignalAntmDiscrete} can be extended to a scenario with multiple targets. Let $L$ be the number of targets, and denote the reflective factor, the spatial frequency, and the delay of the $l$th target by $\alpha_{l}$, ${\vartheta_{l}}$, and $\tau_{l}$, respectively. The reflected echoes from multiple targets are expressed as
	\vspace{-0.1cm} 
	\begin{equation}
	y_{m}^{\left(r\right)}\left[\dtidx\right] = \sum_{l=0}^{L-1}\alpha_{l} h_m\left[\dtidx , \tau_l, {\vartheta_l}\right] + w_{m}^{\left(r\right)}\left[\dtidx\right].
	\label{eqn:MultiRecSignal}
	\vspace{-0.1cm}
	\end{equation}	
	\textcolor{black}{In radar detection, the task is to recover the target parameters  
		$\left\{\tau_{l}, {\vartheta_{l}}, \alpha_{l}\right\}_{l=0}^{L-1}$  from the received signal in \eqref{eqn:MultiRecSignal}.}
	
	\subsubsection{Radar Detection} 
	\label{subsubsec:RadarDetection}
	\ReviseRadar{ The \ac{spacora} radar receiver estimates the parameters $\left\{\tau_{l}, {\vartheta_{l}}, \alpha_{l}\right\}_{l=0}^{L-1}$  using sparse recovery methods.
		To formulate the radar detection scheme, we consider the case in which the target delay $\tau$ satisfies $\tau \in  [\tau_{\rm min}, \tau_{\rm max})$, where $\tau_{\rm min}$ and $\tau_{\rm max}$ are a-priori known minimal and maximal delays, respectively. To recover the parameters of the targets, we divide the range of target delay \textcolor{black}{into a grid of $P$ uniformly-spaced points}  with an interval $\delta \tau \leq \frac{1}{B_{r}}$, where $B_{r}$ is the bandwidth of the radar waveform. Similarly, the spatial frequency range ${\vartheta} \in [-\pi,   \pi)$ is divided into $Q$ equally-spaced points with an interval $\delta \vartheta \leq \frac{2 \pi}{M}$. The grid sets of the discretized delay and spatial frequency are denoted as $\mySet{D} := \left\{\tau^{p} = \tau_{\rm min} + \frac{p}{P}\left(\tau_{\rm max} - \tau_{\rm min}\right)| p = 0, 1, \cdots, P -1 \right\}$, and  
		$\Theta := \left\{{\vartheta}^{q} = -\pi + 2\pi\frac{q}{Q}| q = 0, 1, \cdots, Q - 1\right\}$, respectively. Assuming that the targets are located on the discretized grids, the echoes in \eqref{eqn:MultiRecSignal} can be written as 
		\vspace{-0.1cm}
		\begin{equation}
		\myVec{y}^{\left(r\right)} = \myMat{A}\myVec{b} + \myVec{w}^{\left(r\right)},
		\vspace{-0.1cm}
		\label{eqn:ObserEqn}
		\end{equation}
		where $\myVec{y}^{\left(r\right)}\in \mySet{C}^{N_{\mathrm{rec}}M}$ is the received sample vector whose entries are $\left[\myVec{y}^{\left(r\right)}\right]_{mN_{\mathrm{rec}} + n} := y^{\left(r\right)}_{m}\left[n\right]$; $\myMat{A}$ is  the observation matrix, the entries of which are given by 
		\vspace{-0.1cm}
		\begin{equation}
		\left[\myMat{A}\right]_{mN_{\mathrm{rec}} + n, pQ + q} := h_{m}\left[n, \tau^{p}, {\vartheta}^{q}\right],
		\label{eqn:SensingMat}
		\vspace{-0.1cm}
		\end{equation} 
		and $\myVec{b} \in \mySet{C}^{PQ}$ is a vector with $L$ nonzero entries encapsulating the parameters of the targets.  The $\left(pQ +q\right)$th entry of $\myVec{b}$  equals $\alpha_{l}$ if $\tau^{p} = \tau_l$ and ${\vartheta}^{p} = {\vartheta_{l}}$, while the other entries equal $0$. Here, we assume $\myVec{b}$ is sparse, i.e., $L \ll PQ$. The vector $\myVec{w}^{\left(r\right)}$ is a zero mean white Gaussian noise vector, obtained by $\left[\myVec{w}^{\left(r\right)}\right]_{mN_{\mathrm{rec}} + n} := \myVec{w}_{m}^{\left(r\right)}\left[n\right]$, with variance  $\sigma_{r}^{2}$.}
	
	\ReviseRadar{Due to the sparsity of $\myVec{b}$, it can be recovered by solving 
		\vspace{-0.1cm}
		\begin{equation}
		\min\limits_{\myVec{b}} {\Vert \myVec{b} \Vert}_{0},\  \mathrm{subject\ to}\ {\Vert {\myVec{y}^{\left(r\right)} - \myMat{A}\myVec{b}} \Vert}_{2} \le \epsilon,
		\vspace{-0.1cm}
		\label{eqn:RadarEquCS}
		\end{equation}	
		where $\epsilon$ is related to the noise level.
		The optimization problem \eqref{eqn:RadarEquCS} can be solved by \ac{cs} algorithms, such as greedy approaches, $l_{1}$ norm optimization, and other sparse recovery schemes, as detailed in \cite{eldar2012compressed, eldar2015sampling}. }
	

	\vspace{-0.2cm}
	\section{Radar Performance Analysis}
	\label{sec:RadarPerformance}
	\vspace{-0.1cm}
	\ReviseRadar{In this section we analyze the radar performance of \ac{spacora}. We focus our analysis on the  two-dimensional delay-direction transmit beam pattern, defined as the correlation between the echoes from the steered beam direction and the echoes from the target direction. This measure can be used to characterize the radar resolution and the mutual interference between multiple targets\cite{hu2014randomized}. 
		As the spatial allocation pattern of the transmit array in \ac{spacora} is determined by the transmitted message, the transmit beam pattern of its radar subsystem varies between different communication symbols, thus we compute it by taking the correlation over all symbols transmitted within a single pulse.} 
	%
	In particular, we first characterize the delay-direction transmit beam pattern which arises from the received signal model. Then, we study it statistical properties and compare them to those achieved when using the complete antenna array, as well as with fixed antenna allocation. 
	
	\vspace{-0.2cm}
	\subsection{Transmit Beam Pattern}
	\label{subsec:BeamPattern}
	\vspace{-0.1cm}	

	\ReviseRadar{In the following we characterize the transmit beam pattern of the radar subsystem. Without loss of generality, we express the transmit beam pattern by focusing on the echo captured in a single receive antenna, and particularly on that of index $m=0$. To properly define the transmit beam pattern, we assume that a target  with unit reflective factor is located at   $\left(\tau, {\vartheta}\right)$. The noiseless echo at received antenna $m=0$, i.e., when the noise term in \eqref{eqn:RxEcho1} is nullified,  is given by 
		$y^{\left(r\right)}_{0}\left[n, \tau, \vartheta\right] := h_{0}\left[n, \tau, {\vartheta}\right]$. 
		Similarly, the echo of the  reference target, which is with unit reflective and located in $\left(\tilde{{\tau}}, {\vartheta_{T}}\right)$ is given by  $y^{\left(r\right)}_{0}\left[n, \tilde{{\tau}}, {\vartheta_{T}}\right] $. }
	%
	%
	\ReviseRadar{The transmit delay-direction beam pattern is defined as the correlation of $y^{\left(r\right)}_{0}\left[n, \tau, \vartheta\right]$ and $y^{\left(r\right)}_{0}\left[n, \tilde{{\tau}}, {\vartheta_{T}}\right]$, i.e., $\chi_T\left(\tau,  \vartheta\right) := \sum_{\dtidx = 0}^{\dtmax-1} h_{0}\left[n, \tau, {\vartheta}\right]h_{0}^{\ast}\left[n, \tilde{{\tau}}, {\vartheta_{T}}\right]$. By substituting \eqref{eqn:MatchedFilter} into the definition of $\chi_T\left(\tau,  \vartheta \right)$, we obtain 
		\vspace{-0.1cm} 
		\begin{align}
		\chi_T\left(\tau,  \vartheta\right) &= e^{-j2\pi f_{c}\left(\tau - \tilde{{\tau}}\right)}  \sum_{\dtidx = 0}^{\dtmax-1}\!\sum_{k=0}^{K-1} \tilde{\rho}_{T}   \left(  k, \vartheta  \right) \notag\\
		& \qquad \times  g \left(  \frac{\dtidx T_{s} \!- \!kT_c \! - \! \tau}{T_c}\right) h \left[\dtidx , \tau\right]  h^{\ast}\left[\dtidx , \tilde{\tau}\right].
		\label{eqn:BeampatternTx0}
		\vspace{-0.1cm}
		\end{align}
		The summation over $\dtidx=0,\ldots \dtmax-1$ in \eqref{eqn:BeampatternTx0} represents the averaging of the correlation over the entire radar pulse.
	}

	\ReviseRadar{
		As the phase term $e^{-j2\pi f_c \left(\tau - \tilde{\tau}\right)}$   in \eqref{eqn:BeampatternTx0}  disappears by taking the absolute value, it does not affect the magnitude of the transmit beam pattern. Hence, the term $e^{-j2\pi f_c \left(\tau - \tilde{\tau}\right)}$ is omitted in the sequel, and \eqref{eqn:BeampatternTx0} is rewritten as}
	\vspace{-0.1cm}
	\begin{align}
	\chi_T\left(\tau_{d}, f_{\theta}\right) =  \!\!\!   \sum_{\dtidx = 0}^{\dtmax-1}\!\sum_{k=0}^{K-1}& {\rho}_{T}   \left(  k,f_{\theta}  \right)   g \left(  \frac{\dtidx T_{s} \!- \!kT_c \! - \! \tau_{d} - \tilde{\tau}}{T_c}\right) \notag \\
	& \times  h \left[\dtidx , \tau_{d} + \tilde{{\tau}}\right]  h^{\ast}\left[\dtidx , \tilde{\tau}\right],
	\label{eqn:BeampatternTx}
	\end{align}
	where $\tau_{d} : = \tau - \tilde{\tau}$ is the delay difference, \ReviseRadar{$f_{\theta} : = {\vartheta} - {\vartheta_{T}}$  is the  difference in the  spatial frequency,} and  
	${\rho}_{T}\left(k, f_{\theta} \right) := \tilde{\rho}_{T}  \left( k,\vartheta \right)$. Since the antenna indices $\{m_{k,l}\}$, which are encapsulated in $ \tilde{\rho}_{T}(\cdot, \cdot)$ by \eqref{eqn:RhoT},  are random, it holds that $	\chi_T\left(\tau_{d}, f_{\theta}\right)$ in \eqref{eqn:BeampatternTx} is random. 
	In the following, we analyze the  beam pattern of {\ac{spacora}} compared to using the complete array for radar signaling, 
	as well as to using fixed subsets.

	\vspace{-0.2cm}	
	\subsection{Comparison of Different Antenna Allocation Schemes}	
	\label{subsec:comparison}
	\vspace{-0.1cm}	
	We begin with the transmit beam pattern achieved when utilizing the full antenna array for radar transmission, used as a basis for comparison. 
	Then, the beam patterns of   {\ac{spacora} } as well as fixed antenna allocation methods are evaluated. 
	
	We henceforth focus on radar signalling with chirp waveforms. Here, the baseband radar waveform $h\left(t\right)$ is  
	\vspace{-0.1cm}
	\begin{equation}
	h\left(t\right) = g\left(\frac{t}{T_r}\right)\exp\left\{j\mu \pi \left(t- \frac{T_r}{2}\right)^2 \right\},
	\vspace{-0.1cm}
	\label{eqn:Chirp1}
	\end{equation}
	where $\mu$ is referred to as the frequency modulation rate. The bandwidth of the chirp is defined as $B_r := \mu T_r$. 
	
	\begin{figure}	
		\begin{minipage}[b]{0.48\linewidth}
			\centering
			\centerline{\includegraphics[width=\columnwidth]{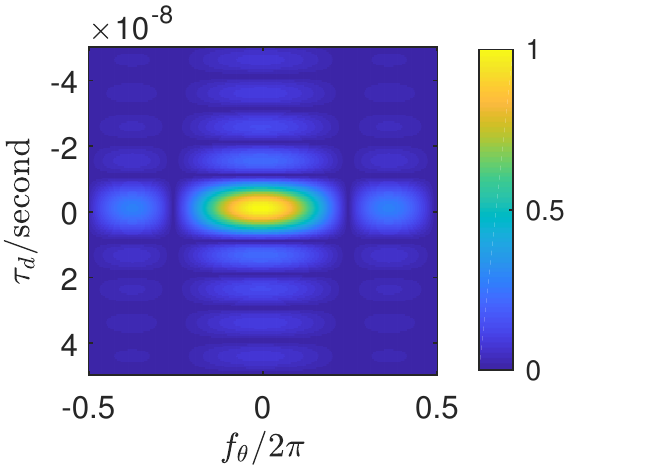}}
			\vspace{-0.1cm}
			\centerline{\small (a) Full Antenna Array}\medskip
		\end{minipage}
		\begin{minipage}[b]{0.48\linewidth}
			\centering
			\centerline{\includegraphics[width=\columnwidth]{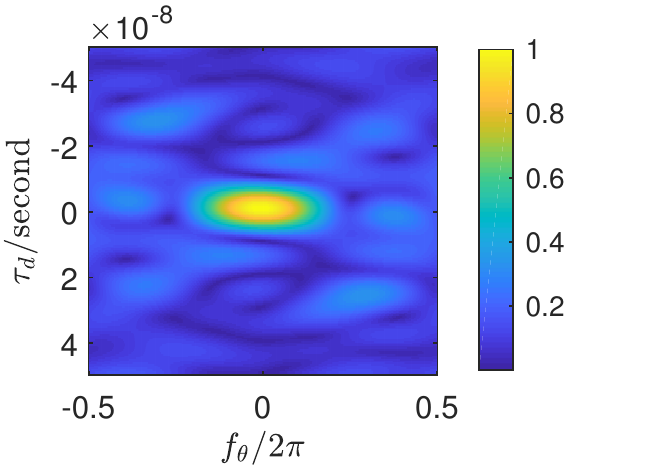}}
			\vspace{-0.1cm}
			\centerline{\small(b) \ReviseDing{\ac{spacora} }}\medskip
		\end{minipage}
		\vspace{-0.1cm}
		
		\begin{minipage}[b]{0.48\linewidth}
			\centering
			\centerline{\includegraphics[width=\columnwidth]{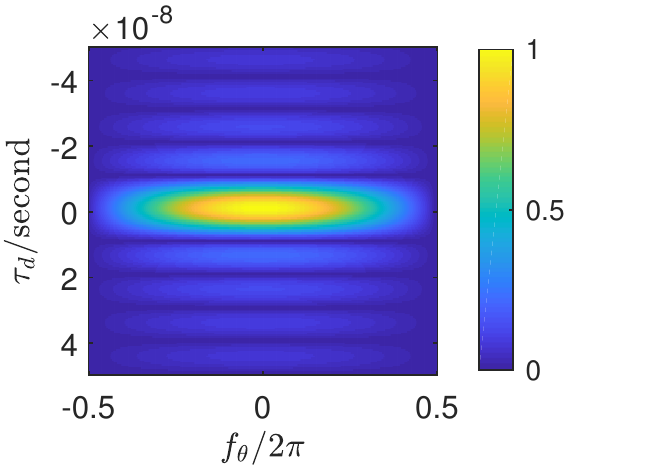}}
			\vspace{-0.1cm}
			\centerline{\small(c) Fix1}\medskip
		\end{minipage}
		\begin{minipage}[b]{0.48\linewidth}
			\centering
			\centerline{\includegraphics[width=\columnwidth]{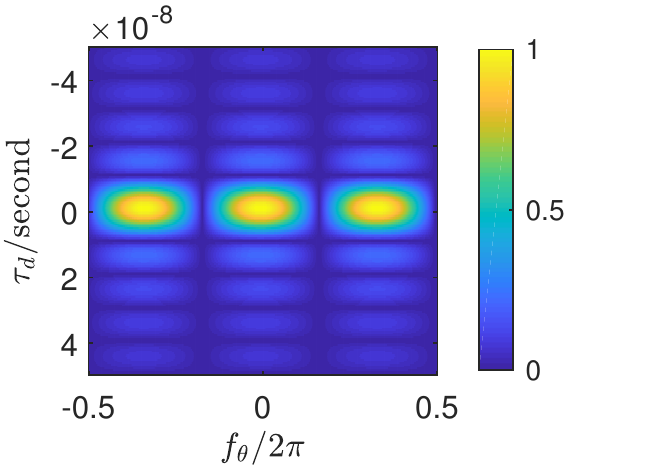}}
			\vspace{-0.1cm}
			\centerline{\small(d) Fix2}\medskip
		\end{minipage}	
		\vspace{-0.3cm}
		\caption{Normalized transmit beam patterns of the analyzed schemes. In this illustration, the parameter are set following Table. \ref{tab:ParameterValue}.
		}
		\label{fig:BeamPatterns}
	\end{figure} 
	
	\subsubsection{Full Antenna Array}
	When the full antenna array is used, the transmit beam pattern can be obtained as a special case of \eqref{eqn:BeampatternTx} by setting $M_{T}^{(r)} = M$. Hence, the antenna indices are deterministic and are given by $m_{k,l} = l$ for each $k = 0, 1, \cdots, K-1$. 
	The resulting transmit delay-direction beam pattern  is a deterministic quantity. 
	
	The full array transmit beam pattern, obtained by substituting \eqref{eqn:Chirp1} and  $M_{T}^{(r)} = M$ into \eqref{eqn:BeampatternTx}, is \cite[Ch. 3A]{cumming2005digital}
	\ReviseDing{
		\begin{equation}
		\left|\chi^{\rm Full}_{T}\left(\tau_{d}, f_{\theta}\right)\right| = \dtrad \left|\mathrm{sinc}\left(B_r \tau_{d} \right)\right|\cdot\left| \frac{\sin\left(Mf_{\theta}/2\right)}{\sin{\left(f_{\theta}/2\right)}}\right|,
		\label{eqn:Beampatternfull}
		\end{equation}	
		where $\dtrad:= \lfloor T_{r}/T_{s}\rfloor$.} 
	The normalized transmit beam pattern for the full antenna array is depicted in Fig. \ref{fig:BeamPatterns}(a). In this beam pattern, 
	the peak located in $\tau_{d} = 0$ and $f_{\theta} = 0$ is denoted as the mainlobe and the other peaks are denoted as the sidelobes. The width of the mainlobe determines the resolution of the radar system, while the sidelobes influence the interference induced by clutters in the environment and the coupling between nearby targets.
	
	\subsubsection{\ReviseDing{\ac{spacora}}}
	In  \ReviseDing{{\ac{spacora} }}, for a given time slot $k$, different indices of the radar transmitting antennas $\{m_{k,l}\}$ are selected. 	As the switching of transmit antennas is determined by the random communication data stream, the transmit beam pattern is a random quantity. 
	\ReviseRadar{The usage of stochastic beam patterns due to random array configuration is an established concept in the radar literature, see, e.g., \cite{lo1964mathematical}. In particular, the following analysis extends the study of  delay-direction beam patterns due to the  randomized switch antenna array with a single active element, investigated in \cite{hu2014randomized}, to multiple active elements.} 
	
	We analyze the statistical moments of the beam pattern, which provide means for evaluating the resolution and the sidelobe level of  \ReviseDing{{\ac{spacora}}}. 
	In particular, we show that the expected beam pattern of \ReviseDing{{\ac{spacora}}}, which is approached by the averaged beam pattern over a large number of pulses, is identical to that of the full antenna array up to a constant factor. 
	%
	%
	%
	%
	%
	This holds due to the following theorem:
	\begin{theorem}
		\label{thm:ExpBeamPattern}
		The absolute value of the expected transmit delay-direction beam pattern \eqref{eqn:BeampatternTx} of { \ac{spacora}} is 
		\begin{align} 
		&\left|\mathcal{E}\left\{\chi_{T}^{\rm GSM}\left({\tau_{d}},f_{\theta}\right)\right\}\right|\notag\\
		&=\!\frac{M_{T}^{r}}{M}\! \ReviseDing{\left|\sum_{\dtidx=0}^{\dtmax-1}h\!\left[\dtidx , \tau_{d} + \tilde{{\tau}}\right] \!h^{\ast}\left[\dtidx , \tilde{\tau}\right]\right|}\!\cdot\!\left|\frac{\sin\!\left(Mf_{\theta}/2\right)}{\sin\!{\left(f_{\theta}/2\right)}}\right|. 
		\label{eqn:ExpBeamPattern}
		\end{align}
	\end{theorem}
	
	\begin{IEEEproof}
		The proof is given in Appendix  \ref{app:Proof1}. 	
	\end{IEEEproof}
	
	The expectation in \eqref{eqn:ExpBeamPattern} is carried out with respect to the random antenna  indices $\{m_{k,l}\}$. These indices are determined by the communicated bits, which are assumed to be i.i.d.. It follows from the law of large number that as the number of pulses grows, 
	the average transmit beam pattern approaches its expected value with probability one \cite[Ch. 8.4]{papoulis2002probability}.
	Consequently, in the  large number of pulses horizon, the magnitude of the average transmit beam pattern coincides with \eqref{eqn:ExpBeamPattern}. 
	
	Theorem \ref{thm:ExpBeamPattern} is formulated for arbitrary waveforms $h(t)$. 
	For chirp signals, it is specialized in the following corollary:
	\begin{corollary}
		\label{cor:ExpBeamP}
		The absolute value of the expected transmit  beam pattern \eqref{eqn:BeampatternTx} for  \ReviseDing{{ \ac{spacora} }} with chirp waveform is 
		\begin{align} 
		\small
		\!\!\!\left|\mathcal{E} \! \left\{\chi_{T}^{\rm GSM} \! \left(\tau_{d},f_{\theta}\right)\right\} \! \right|\! = \! \frac{M_{T}^{r}\dtrad}{M}\!\left| \mathrm{sinc} \! \left( \! B_r \tau_{d} \!\right)  \right| \!\! \cdot \!\! \left| \!  \frac{\sin\left(Mf_{\theta}/2\right)}{\sin{\left(f_{\theta}/2\right)}} \! \right|.
		\label{eqn:ExpBeamPattern_chirp}
		\end{align}
	\end{corollary}
	
	\begin{IEEEproof}
		The corollary is obtained by substituting \eqref{eqn:Chirp1} into the expected transmit beam pattern in \eqref{eqn:ExpBeamPattern}.
	\end{IEEEproof}
	
	Corollary \ref{cor:ExpBeamP} implies that   \ReviseDing{ \ac{spacora} }, which utilizes the antenna array for both radar signalling and communication transmission without using multiband signals, has the same expected beam pattern as in \eqref{eqn:Beampatternfull} (up to a constant factor), i.e., the same as when using the complete  array only for radar.  
	This implies that, e.g., when averaged over a large number of pulses,  \ReviseDing{{ \ac{spacora} }}   achieves the same ratio of the sidelobe level to the mainlobe as that of using the complete  array for radar.
	
	
	
	For a single radar pulse with a finite number of symbols, the difference between the (random) instantaneous transmit beam pattern and its expected value is dictated by its variance \cite[Ch. 5]{papoulis2002probability}. Consequently, larger variance induces increased fluctuations in the transmit beam patterns compared to its expected value \eqref{eqn:ExpBeamPattern_chirp}. The variance of the transmit beam pattern with chirp waveforms is stated in the following proposition:
	\begin{proposition}
		\label{pro:Variance}
		The variance of the normalized transmit delay-direction beam pattern \eqref{eqn:BeampatternTx} with chirp waveform \eqref{eqn:Chirp1} is
		\vspace{-0.1cm} 
		\begin{equation}
		\begin{aligned} 
		&\mathcal{V}\left\{\chi_{T}^{\rm GSM}\left(\tau_{d}, f_{\theta}\right)/\mathcal{E} \left\{\chi_{T}^{\rm GSM}\left( 0, 0\right)\right\}\right\} = \gamma^{\rm GSM}\left(\tau_{d}\right)\notag \\
		& \times\left[ \! \frac{\left(M_{T}^{r}\!-\! M\right)}{M_{T}^{r}M^2\left(M-1\right)} \!\!\cdot\!\! \left|\frac{\sin\left(Mf_{\theta}/2\right)}{\sin\left(f_{\theta}/2\right)}\right|^2 \!\!\!+\! \frac{\left(M\!-\! M_{T}^{r}\right)}{M_{T}^{r}\left(M-1\right)} \! \right], 
		\label{eqn:VarianceBeamPattern}
		\end{aligned}
		\vspace{-0.1cm}
		\end{equation}
		where 
		\vspace{-0.1cm}
		\begin{equation}
		\!\gamma^{\rm GSM}\!\left(\tau_{d}\right) \!:=\! \frac{\mathrm{sinc}^2\left({B_r\tau_{d}}/{K}\right)}{K}\! = \! \frac{\mathrm{sinc}^2\!\left(\mu T_c\tau_{d}\right)}{K}.
		\label{eqn:VarGamma}
		\end{equation}
	\end{proposition}
	
	\begin{IEEEproof}
		The proof is given in Appendix \ref{app:Proof2}.	
	\end{IEEEproof}
	
	From Proposition \ref{pro:Variance} it follows  that the  variance decreases when $K$  increases. A small variance leads to an improved beam pattern, as it is less likely to deviate from its desired mean value when the variance of the beam pattern decreases.	
	
	The similarity between the  beam patterns of \ReviseDing{{ \ac{spacora}}} and that of using the full array is demonstrated in Fig.~\ref{fig:BeamPatterns}. In particular, Fig. \ref{fig:BeamPatterns}(b) is a realization of the average beam pattern in a single pulse of   \ReviseDing{{ \ac{spacora} }} with $K = 12$ symbols and $M_{T}^{(r)} = 2$ antennas assigned for radar, while Fig. \ref{fig:BeamPatterns}(a) is the corresponding beam pattern when using all the $M = 4$ elements for radar.  
	Comparing Fig. \ref{fig:BeamPatterns}(a) and Fig. \ref{fig:BeamPatterns}(b) shows the similarity of the beam pattern of \ReviseDing{ \ac{spacora}},  in terms of mainlobe width and sidelobe levels, to that achieved when using the full antenna array for radar.
	
	\subsubsection{Fix1 Scheme}
	In this scheme, the full antenna array is divided into two sub-\acp{ula}, one for radar and one for communication, in a fixed manner. The indices of the transmit  elements are  $m_{k,l} = l$ for each $k$, as when using the full array for radar signalling. However, here only a subset of the array is used for radar, i.e., $M_{T}^{(r)} < M$.  
	The resulting deterministic transmit  beam pattern is stated in the following proposition:
	\begin{proposition}
		\label{pro:BeamPFix1}
		The transmit beam pattern of {\em Fix1 } with chirp waveforms is given by
		\begin{align}
		\left|\chi^{\rm Fix1}_{T} \! \left(\tau_{d}, f_{\theta}\right)\right| \!\!=\!\! \frac{M_{T}^{r}\dtrad}{M} \!\left|\mathrm{sinc}\left(B_r \tau_{d} \right)\right| \!\cdot\! \left| \frac{\sin\left(M_{T}^{r}f_{\theta}/2\right)}{\sin{\left(f_{\theta}/2\right)}}\right|.
		\label{eqn:BeamPFix1}
		\end{align}
	\end{proposition}
	
	\begin{IEEEproof}
		The proposition is obtained by substituting the  transmit antenna indices $m_{k,l} = l$
		and \eqref{eqn:Chirp1} into \eqref{eqn:BeampatternTx}. 
	\end{IEEEproof} 
	
	As $M_{T}^{r} < M$, the mainlobe in \eqref{eqn:BeamPFix1} is wider then that of  \eqref{eqn:Beampatternfull}. 
	The normalized transmit beam pattern of \emph{Fix1} is depicted in Fig. \ref{fig:BeamPatterns}(c), which is computed using the same settings as in Figs. \ref{fig:BeamPatterns}(a)-\ref{fig:BeamPatterns}(b), where it is indeed observed that its mainlobe is wider than that of the full antenna array.
	
	\subsubsection{Fix2 Scheme}	
	An alternative  allocation approach is to randomly divide the antenna array into two sub-arrays: One for radar and the other for communications. In this method,  the indices of the transmit elements are randomized and remain unchanged during the whole radar pulse duration, i.e., $\{m_{k,l}\}$ is the same set of random variables for each $k=0,1,\ldots,K-1$. A realization of the normalized transmit beam pattern for \emph{Fix2} is depicted in Fig. \ref{fig:BeamPatterns}(d). When we only consider the radar subsystem, this approach can be regarded as a specific case of \ac{spacora} by setting $K = 1$, and the expected value and variance of the transmit delay-direction beam pattern are obtained by substituting $K = 1$ into \eqref{eqn:ExpBeamPattern} and \eqref{eqn:VarianceBeamPattern}, respectively. As the parameter $K$ does not affect the expectation of the transmit beam pattern \eqref{eqn:ExpBeamPattern_chirp}, it holds that the expected transmit delay-direction beam pattern of \emph{Fix2}  is the same as that of \ac{spacora}. However \emph{Fix2} has a higher sidelobe level compared with \ac{spacora}, which can be evaluated through its variance, as stated in the following corollary:
	\begin{corollary}
		The variance of the normalized transmit delay-direction beam pattern is written as 
		\begin{align} 
		&\mathcal{V}\left\{\chi_{T}^{\rm Fix2}\left(\tau_{d}, f_{\theta}\right)/\mathcal{E}\left\{\chi_{T}^{\rm Fix2}\left(0,0\right)\right\}\right\} = \gamma^{\rm Fix2} \left(\tau_{d}\right)\notag \\
		&\times \left[\!\frac{ \!\left(M_{T}^{r} \! - \!M\right)}{M_{T}^{r}M^2\!\left(M \! - \!1\right)} \!\cdot\! \left|\frac{\sin\left(Mf_{\theta}/2\right)}{\sin\left(f_{\theta}/2\right)}\right|^2 \!\!+\! \frac{\!\left( \!M \!-\! M_{T}^{r} \!\right)}{M_{T}^{r}\left(M\!-\!1\right)}\!\right], 
		\label{eqn:VarianceBeamPattern_c}
		\end{align}
		where
		\begin{equation}
		\gamma^{\rm Fix2}\left(\tau_{d}\right) := \mathrm{sinc}^2\left(B_r\tau_{d}\right).
		\label{eqn:VarGamma_c}
		\end{equation}
	\end{corollary}
	
	\begin{IEEEproof}
		Setting $K = 1$ in Proposition \ref{pro:Variance} proves \eqref{eqn:VarianceBeamPattern_c}.
	\end{IEEEproof}
	
	Comparing \eqref{eqn:VarGamma} with \eqref{eqn:VarGamma_c}, we find that for a given pulse width $T_r$, the maximal variance of the transmit beam pattern for \emph{Fix2}, i.e., \eqref{eqn:VarGamma_c} for $\tau_d = 0$, is $K$ times that of \ReviseDing{{ \ac{spacora}}}. This demonstrates that the dynamic changing of antenna elements, whose purpose in \ReviseDing{{ \ac{spacora}}} is to increase the communications rate, allowing to convey more \ac{gsm} symbols in each radar pulse, also improves the radar angular resolution  and decreases the sidelobe levels. 
	%
	The performance advantages of \ReviseDing{{ \ac{spacora}} over the fixed antenna allocation approaches} are numerically observed in Section \ref{sec:Sims}.

	\section{Hardware Prototype High Level Design}
	\label{sec:Highleveldesign}
	\vspace{-0.1cm}
	To demonstrate the feasibility of the proposed \ac{dfrc} system, we implemented \ac{spacora} using a dedicated hardware prototype. 
	\ReviseDing{This prototype, used here to experiment \ac{spacora}, can realize a multitude of \ac{dfrc} systems, as it allows baseband waveform generation, over-the-air signaling, frequency band waveform transmission, radar echo generation, radar echo reception, and communication signal reception.} 
	In this section, we describe the high level design of the prototype, detailing the structure of each component  in Section \ref{sec:ProtoRealization}. The overall system structure is described in Subsection \ref{subsec:Highlevelstructure}, and in Subsection \ref{subsec:SystemPara}, we introduce how to choose the system parameters. Finally, in Subsection \ref{subsec:WaveformGenerator}  we present how the \ac{jrc} waveforms transmitted by each antenna element are generated in the prototype.  
	
	\vspace{-0.2cm}
	\subsection{Overall System Architecture}
	\label{subsec:Highlevelstructure}
\begin{figure}
		
		\begin{minipage}[b]{1.0\linewidth}
			\centerline{\includegraphics[width=\columnwidth]{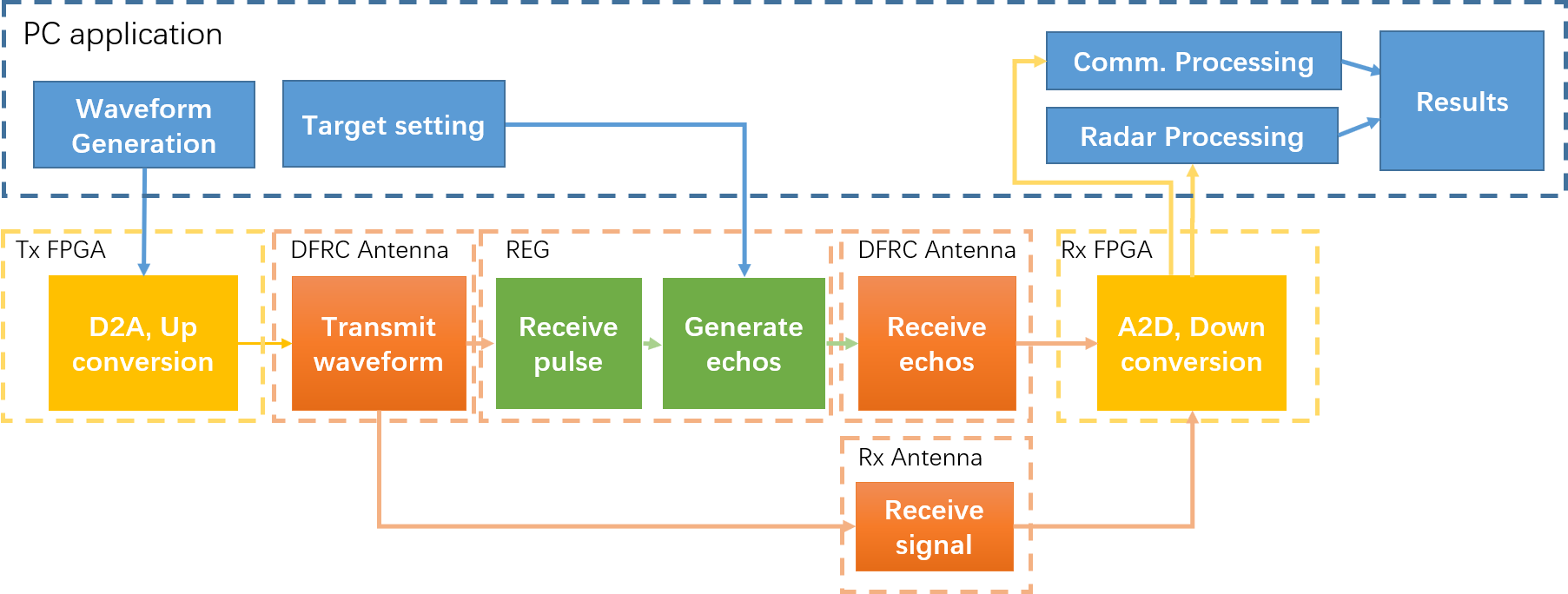}}
			\centerline{\small (a) Experimental setup flow diagram}\medskip
		\end{minipage}
		
		\begin{minipage}[b]{1.0\linewidth}
			\centerline{\includegraphics[width=\columnwidth]{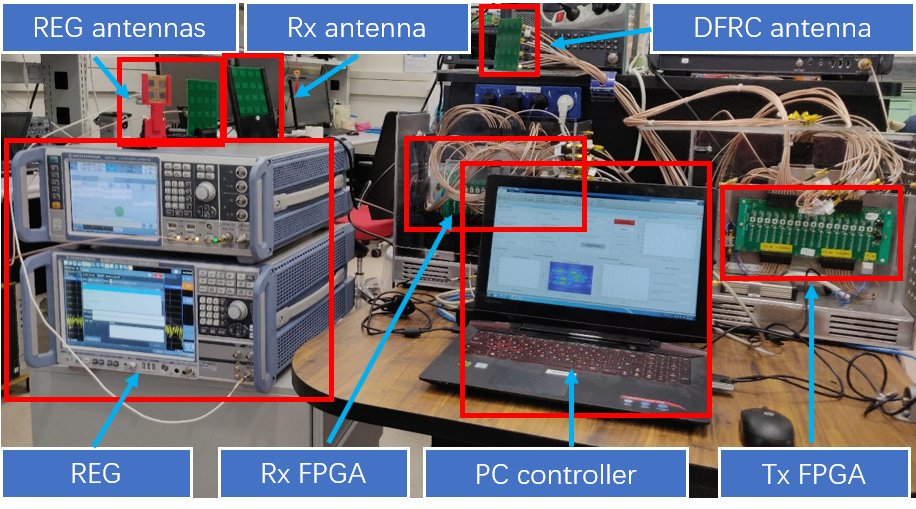}}
			\centerline{\small (b) \ac{dfrc} prototype}\medskip
		\end{minipage}
		\vspace{-0.6cm}
		\caption{The high level structure and components of the \ac{dfrc} prototype.}
		\label{fig:Highleveldesign}
	\end{figure}

	\vspace{-0.1cm}
	The overall structure of the prototype and the high level information flow of the experimental setup are depicted in Fig. \ref{fig:Highleveldesign}. Our setup consists of
	$1)$ a PC server, which provides \ac{gui} for setting the \ac{dfrc} parameters, generates the waveforms, and processes the received signals; $2)$ \ReviseDing{a two-dimensional digital antenna array with $16$ elements, which enables to independently control each element. In our experiment, the array  is divided into $8$ transmit elements and $8$ receive elements;} $3)$ a pair of \ac{fpga} boards interfacing the \ac{dfrc} transmitted and received signals, respectively, between the  PC and the antenna; and $4)$ a \ac{reg} which receives the transmitted  waveform and generates the reflected echoes.  
	
	Through the \ac{gui}, the parameters of the radar and communication subsystems, as well as those of the experimental setup, are configured. Once the paramerers are set and an experiment is launched, the \ac{jrc} waveform is generated by the PC application. Then, the \ac{jrc} waveform is transferred to the \ac{dfrc} transmit \ac{fpga} in which it is converted into analog, up-converted to passband, and forwarded to the \ac{dfrc} antenna array for transmission.
	The transmitted waveform is received by the \ac{reg}, which in response transmits echoes simulating the presence of radar targets, as well as by a receive antenna, which is connected to the receive \ac{fpga}.
	The received radar echoes at the \ac{dfrc} antenna and the received communications signal are down-converted, digitized and then sent to the PC server. The digitized signals are processed by the PC application, which in turn recovers the radar targets and the communication messages.

	\vspace{-0.2cm}
	\subsection{System Parameterization}
	\label{subsec:SystemPara}
	\vspace{-0.1cm}
	In order to guarantee the performances of both radar and communications systems, several criteria should be considered when designing the system parameters. The design of radar waveform parameters, including \ac{pri}, radar bandwidth, pulse width, etc., have been well studied and can be found in \cite{Skolnik2001Inroduction, Skolnik2008Radar}. Here, we discuss how to choose the parameters unique to the proposed \ac{dfrc} system, i.e., $M_{T}^{r}$, $M_{T}^{c}$ and $K$. 
	
	\subsubsection{Number of Elements Allocated for Radar}
	For the radar subsystem, which is considered to be the primary functionality, the maximal detection range is related to the antenna transmit gain, which approximately equals $\left(M_{T}^{r}\right)^{2}P_{t}$. Hence, the minimum number of antenna elements should satisfy the requirement of radar detection range and can be determined according to the radar equation \cite[Ch. 2]{Skolnik2001Inroduction}. Once $M_{T}^{r}$ is determined, the value of $M_{T}^{c}$ is obtained as $M_{T}^{c} = M - M_{T}^{r}$. 
	
	\subsubsection{Number of Chips Divided}
	Based on the radar performance analysis presented in Section \ref{sec:RadarPerformance}, the variance of the transmit beam pattern decreases with the increase of $K$. This indicates that larger values of $K$ are preferable. Furthermore, for the radar subsystem, the chirp is divided into $K$ short chips. The bandwidth with a rectangular window function  of  duration $T_{r}/K = T_c$ is $1/Tc$, which is the bandwidth of the communications signal. If $1/T_c$ is larger than $B_{r}/K$, the bandwidth of a short chirp chip will be expanded in frequency spectrum. Thus, we require $B_{r}/K > 1/T_c = K/T_{r}$, i.e, $K^{2} < B_{r}T_{r}$. 
	Finally, the communication channel is assumed to be flat, requiring the bandwidth of the communications signal to be smaller than the coherence bandwidth, i.e., $1/T_{c} < B_{c}$, where $B_{c}$ is the coherence bandwidth of the communication channel. {To summarize, in light of the aforementioned considerations, the value of $K$ should be set to $K < \min\left\{\sqrt{B_{r}T_{r}}, T_{r}B_{c}\right\}$}.

	\vspace{-0.2cm}
	\subsection{Generation of the \ac{jrc} Waveform}
	\label{subsec:WaveformGenerator}
	\vspace{-0.1cm}
	As detailed in Section \ref{sec:Model}, our system implements radar and communications  by allocating different antenna elements to each functionality in a randomized fashion. Unlike traditional \ac{gsm} communications in which the active antenna are changed using  switching \cite{Renzo2014Spatial}, our prototype embeds the randomized  allocation pattern into a dedicated \ac{jrc} waveform. Here, we describe how these joint waveforms are generated. 
	
	
	\begin{figure}	
		\begin{minipage}[b]{0.68\linewidth}
			\centering
			\centerline{\includegraphics[width= \columnwidth]{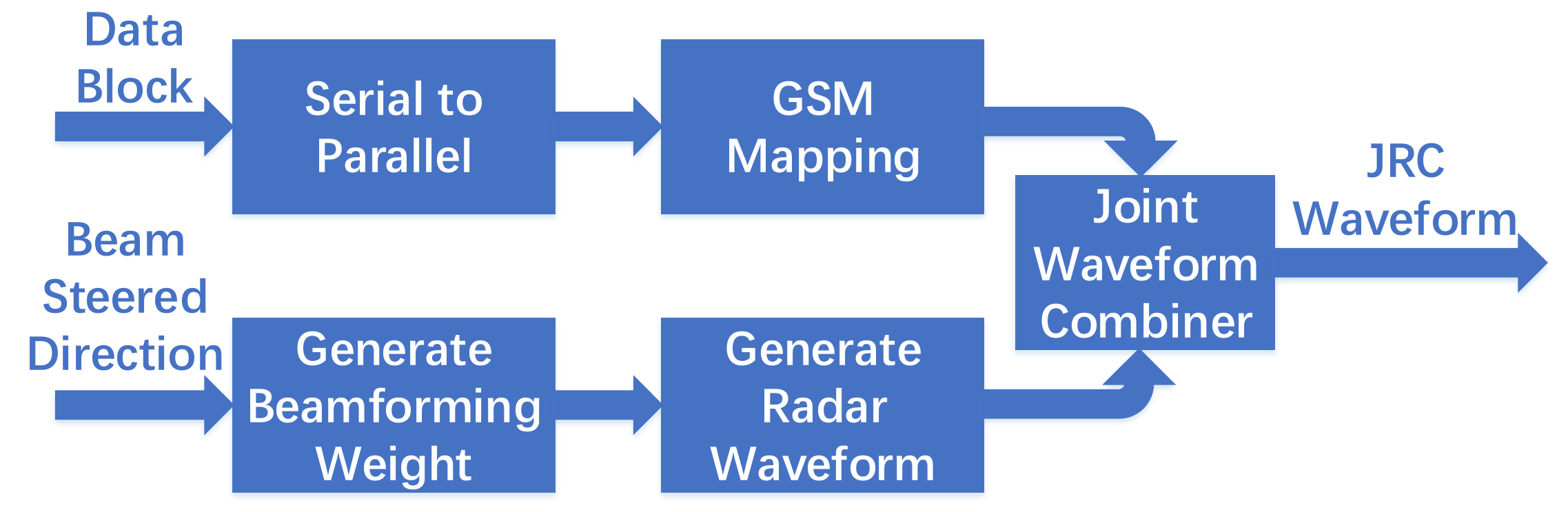}}
			\centerline{(a)}\medskip
		\end{minipage}
		\begin{minipage}[b]{0.28\linewidth}
			\centering
			\centerline{\includegraphics[width=\columnwidth]{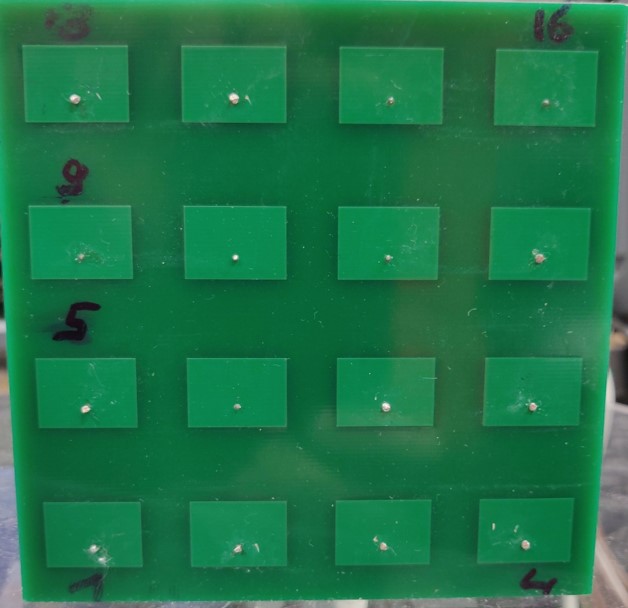}}
			\centerline{(b)}\medskip
		\end{minipage}
		\vspace{-0.4cm}
		\caption{Waveform generator block diagram (left) and antenna array (right).}
		\label{fig:WaveformGenerator}
	\end{figure}

	A block diagram of the \ac{jrc} waveform generator is depicted in Fig. \ref{fig:WaveformGenerator}(a): 
	The inputs of the generator are the communication data block and the steered direction of the radar beam. The output of the waveform generator is the \ac{jrc} waveform for each antenna element. The \ac{jrc} waveform transmitted  determines the allocation pattern   of the antenna array. The generation process consists of the following steps: 
	\begin{enumerate}
		\item Communication symbol generator: the conversion of the data block into \ac{gsm} symbols consists of two modules:
		\begin{itemize}
			\item Serial-to-parallel (S/P) module, where the data block is divided into multiple \ac{gsm} blocks. Each \ac{gsm} block consists of two sets of bits: spatial selection bits, used for determining the antenna allocation, and constellation bits, conveyed in the communication symbol.
			\item \ac{gsm} mapping module, which maps  each \ac{gsm} block into its corresponding constellation symbol  and antenna allocation pattern. 
		\end{itemize}
		\item Radar waveform generator: the beam direction is converted into a radar waveform via the following modules:
		\begin{itemize}
			\item Beamforming weight generation, which assigns the weights to direct towards the steered direction. 
			\item Radar waveform generation, which weights the initial radar waveform to obtain the desired beampattern.
		\end{itemize}
		\item Radar and communications waveform combiner: the \ac{jrc} waveform is generated by combining the radar waveform and communication symbol blocks. In this combiner, the communication chips are inserted into the radar waveform based on the antenna allocation bits. An example for such a combined waveform is depicted in Fig. \ref{fig:AntennaAlocationSchemes}(c).
	\end{enumerate}
	
	As illustrated in  Fig. \ref{fig:AntennaAlocationSchemes}(c), the \ac{jrc} waveform is divided into multiple time slots, where the length of each slot is dictated by the communication symbol duration. 
	In each time slot, the allocation of the array element is determined by the content of its waveform. This joint waveform generation process facilitates the application of \ReviseDing{{\ac{spacora}}} without utilizing complex high speed switching devices.

	\vspace{-0.2cm}
	\section{Prototype Realization}
	\label{sec:ProtoRealization}
	\vspace{-0.1cm}
	In the previous section, we introduced the design philosophy of the \ac{dfrc} prototype, which divides the implementation of our scheme between software and dedicated hardware components. The prototype is depicted in Fig. \ref{fig:Highleveldesign}(b): It consists of a PC server, a \ac{dfrc} Tx board, a \ac{dfrc} Rx board, a \ac{reg}, and a two dimensional antenna array with $16$ elements. 
	The GUI and data processor are implemented in software on the PC server. In this section we present the structure of each component in our \ac{dfrc} prototype, detailing the hardware and software modules in Subsections \ref{subsec:HardwareComponents}-\ref{subsec:GUI}, respectively.
	
	\vspace{-0.2cm}
	\subsection{Hardware Components}
	\label{subsec:HardwareComponents}
	\subsubsection{Antenna Array}
	\label{subsubsec:Antenna}
	A two dimensional antenna array with $16$ elements is used in our prototype, depicted in Fig. \ref{fig:WaveformGenerator}(b). The antenna works with carrier frequency $5.1$ GHz and has a $80$ MHz bandwidth. In particular, we use the frequency band of $5.06-5.11$ GHz for radar, while the band $5.11-5.14$ GHz is assigned to communication.
	%
	%
	The antenna consists of $16$ elements, where $8$ are utilized for transmission and $8$ for receive. The selection of which elements are used is determined by a set of 16 switches. The antenna switching is controlled by a micro-controller with an internal memory which is  controlled by the PC application via serial interface. 
	\subsubsection{\ac{dfrc} Tx Board}
	\label{subsubsec:Transmitter}
	The input of the transmitter is a set of  $8$ digital \ac{jrc} waveforms generated by the PC server, each intended for a different element. In the transmitter, the digital \ac{jrc} waveform is converted into analog,  up-converted to passband, and amplified. The resulting analog waveform is forwarded to the transimit antenna using $8$ cables.
	
	\begin{figure}	
		\begin{minipage}[b]{0.48\linewidth}
			\centering
			\centerline{\includegraphics[width= \columnwidth]{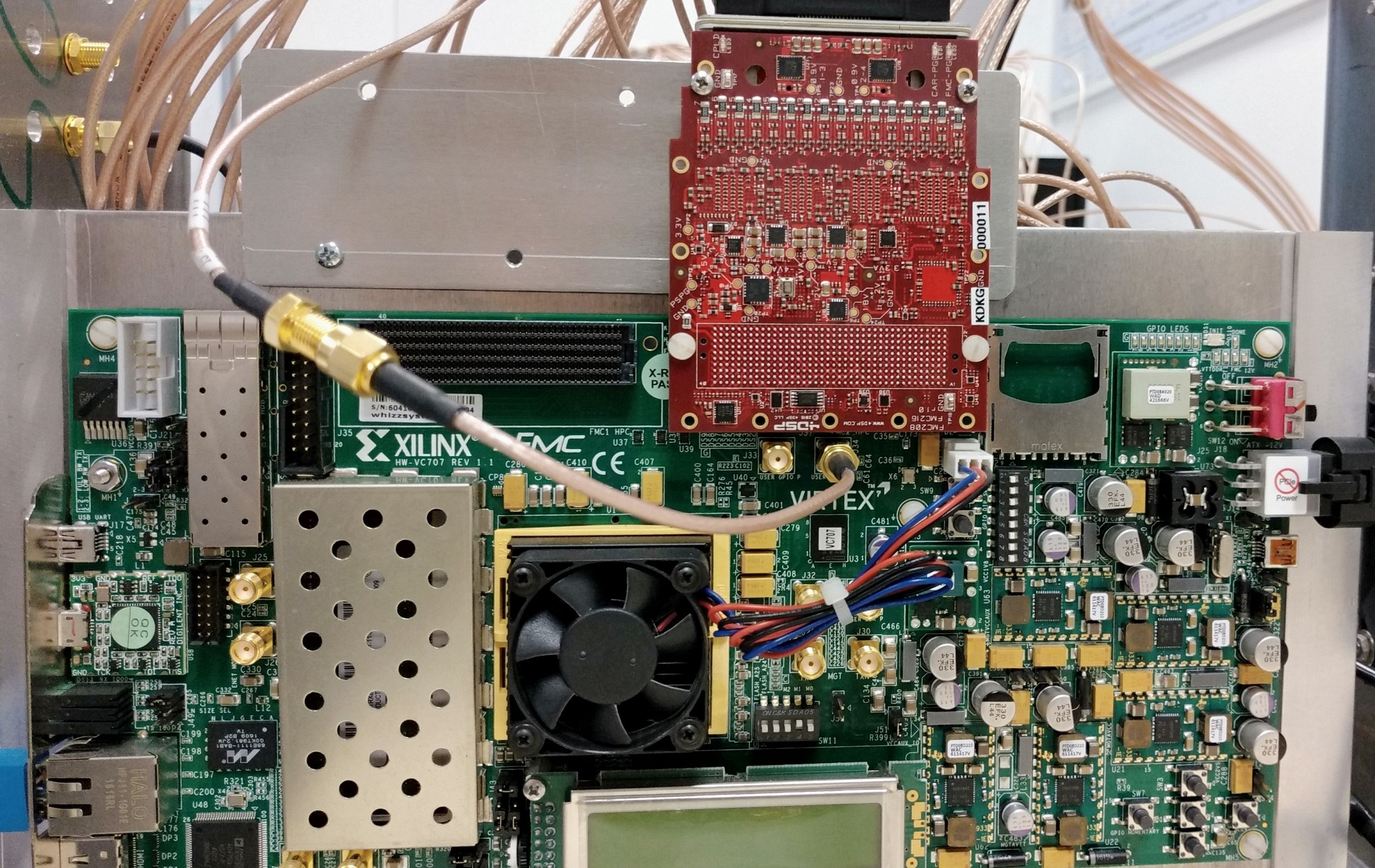}}
			\centerline{(a)}\medskip
		\end{minipage}
		\begin{minipage}[b]{0.48\linewidth}
			\centering
			\centerline{\includegraphics[width=\columnwidth]{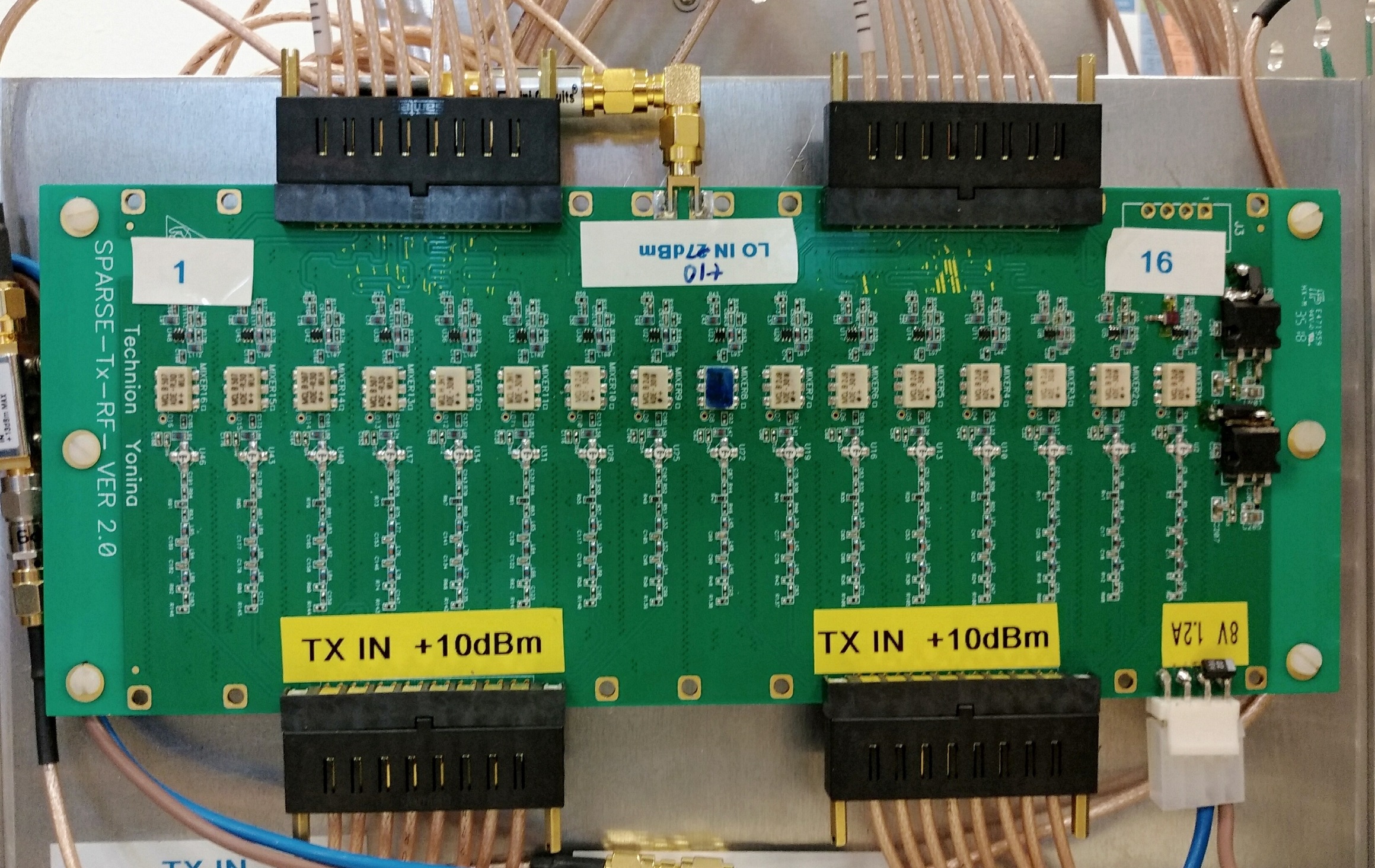}}
			\centerline{(b)}\medskip
		\end{minipage}
		\vspace{-0.4cm}
		\caption{FPGA board, DAC card Radio frequency card of the transmitter.}
		\label{fig:FPGAboardTx}
	\end{figure} 
	
	This process is implemented using three components: An \ac{fpga} board, a \ac{dac} card, and an up-conversion card. These components are depicted in Fig.~\ref{fig:FPGAboardTx}. High speed data transmission interface is realized on the \ac{fpga}, which transfers the digital waveform data from the portable server to the \ac{dac} board. Each of the $8$ digital signals is converted to analog using  a 4DSP FMC216 \ac{dac} card. The FMC216 provides sixteen 16-bit DAC at 312.5Msps (interpolated to 2.5Gsps) based on TI DAC39J84 chip. In the up-conversion card, the analog waveform is up-converted using a local oscillator and amplified by a passband filter. After digital to analog conversion and up conversion, $8$ waveforms are forwarded to the antenna array to be transmitted.
	
	\subsubsection{\ac{dfrc} Rx Board}
	\label{subsubsec:Receiver}
	The receiver board allows the received radar echoes to be processed in software. Broadly speaking, it converts the passband analog echoes and received waveforms to baseband digital streams, forwarded to the server.
	\begin{figure}	
		\begin{minipage}[b]{0.48\linewidth}
			\centering
			\centerline{\includegraphics[width=\columnwidth]{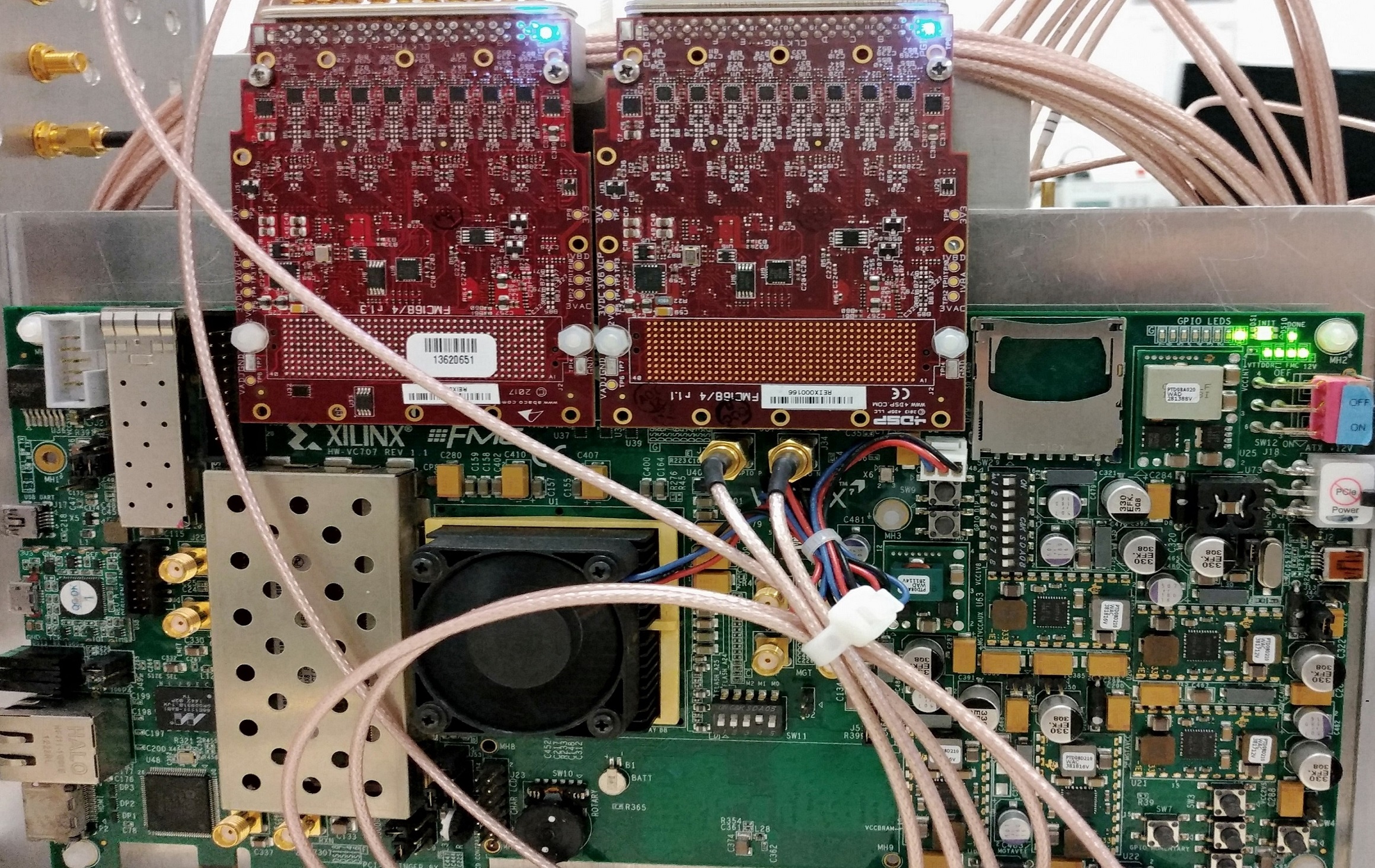}}
			\centerline{(a)}\medskip
		\end{minipage}
		\begin{minipage}[b]{0.48\linewidth}
			\centering
			\centerline{\includegraphics[width=\columnwidth]{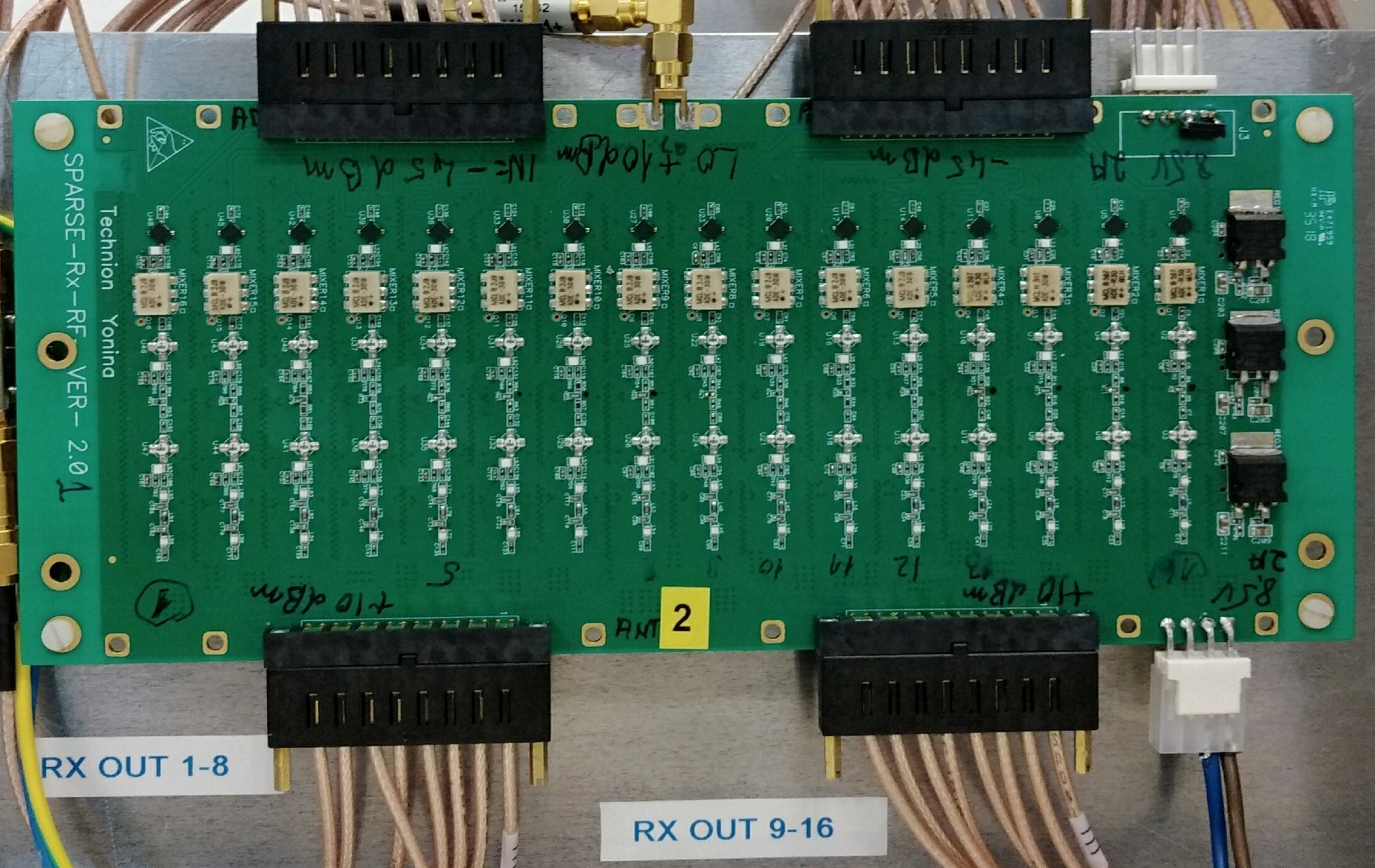}}
			\centerline{(b)}\medskip
		\end{minipage}
		\vspace{-0.4cm}
		\caption{FPGA board, ADC cards and radio frequency card of the receiver.}
		\label{fig:FPGAboardRx}
	\end{figure} 
	The receiver board consists of a VC707 FPGA  board, two FMC168 \acp{adc} cards, and a radio frequency down-convertor board, as depicted in Fig.~\ref{fig:FPGAboardRx}. Each received passband signal is first down converted to baseband by the radio frequency card. This process consists of amplifying the passband waveform, followed by being mixed with a local oscillator, and applying a baseband filter, resulting in a baseband signal, which is in turn  amplified by a baseband amplifier. The amplified analog baseband waveform is  converted to digital by the FMC168 card. The FMC168 is a digitizer featuring $8$ \ac{adc} channels based on the TI ADS42LB69 dual channel 16-bit 250Msps A/D. The board is equipped with two \ac{adc} cards, where one is connected to the receive elements of the \ac{dfrc} antenna and the other is connected to the communications receiver antenna. The high speed data transmission interface is implemented on the \ac{fpga}, transferring the signals to the PC server, where they are processed via the detection strategy detailed in Section \ref{sec:Model}.
	
	\subsubsection{\ac{reg}}
	\label{subsubsec:REG}
\begin{figure}	
		\begin{minipage}[b]{1.0\linewidth}
			\centerline{\includegraphics[width=\columnwidth]{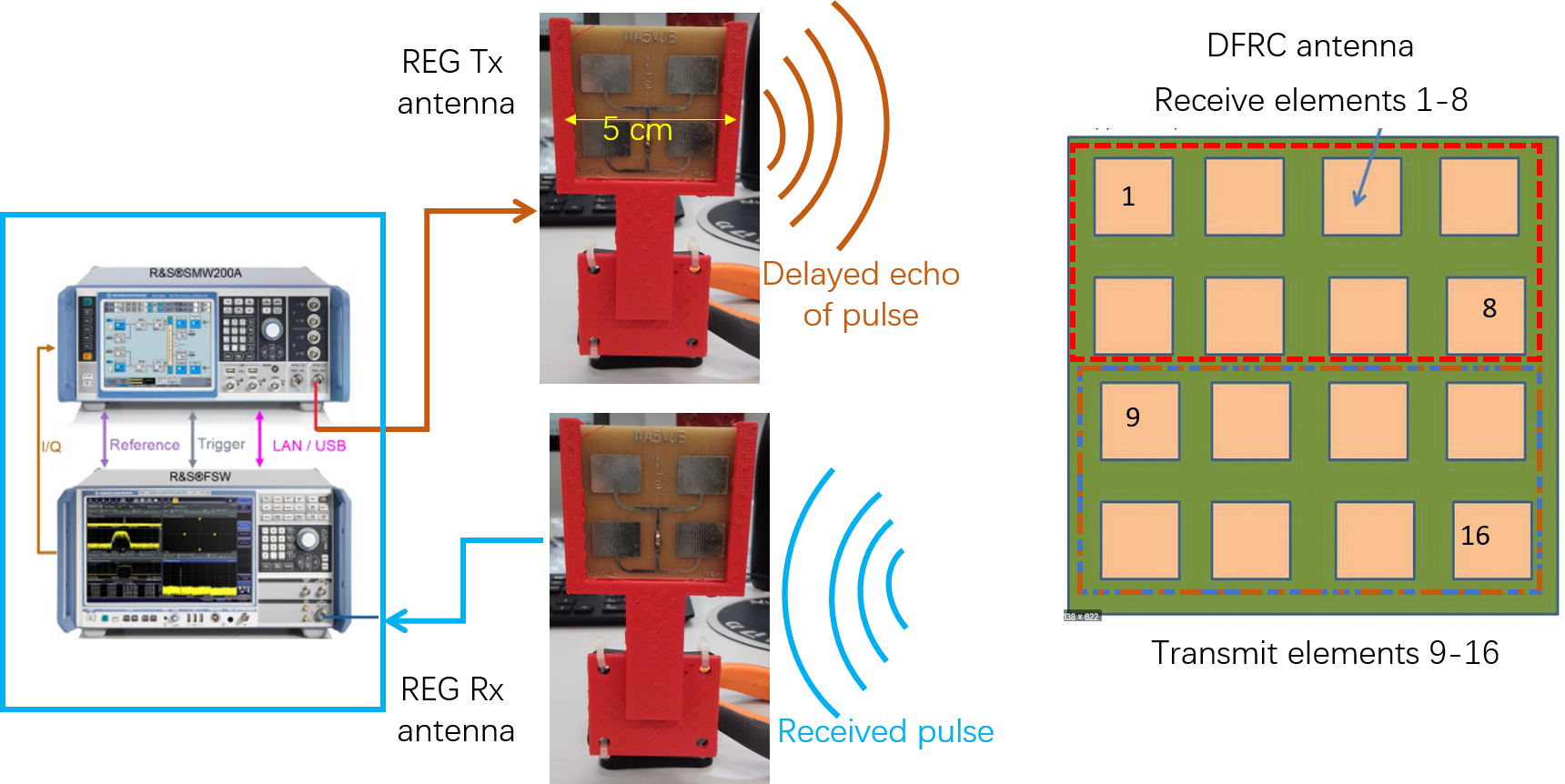}}
		\end{minipage}
		\vspace{-0.4cm}
		\caption{Schematic illustration of \ac{reg} operation.}
		\label{fig:REG}
	\end{figure}

	In order to simulate echoes generated by moving radar targets in an over-the-air setup, we use a \ac{reg}. The \ac{reg} consists of a Rhode \& Schwarz FSW signal and spectrum analyzer, which captures the received waveform, and a Rhode \& Schwarz SWM200A vector signal generator, which adds the delays and Doppler shifts to the observed waveform and transmits it over-the-air. 
	The signal and spectrum analyzer and the vector signal generator are connected to a dedicated receive and transmit antenna element, respectively. An illustration of the \ac{reg} components and their operation is depicted in Fig. \ref{fig:REG}. 
	The \ac{reg} operation is triggered when it receives a transmitted radar pulse.  This procedure allows us to experiment our prototype with over-the-air signaling with controllable targets.
	Up to  $6$ targets can be generated by the \ac{reg}, whose range and Doppler can be configured to up to $10$ kM and $190$ kHz, respectively. The parameters of the targets are configured directly by the PC application by LAN interface.

	\subsubsection{Data Processor}
	\label{subsubsec:Processor}
	The data processor is a 64-bit laptop with 4 CPU cores and a 16GB RAM. A Matlab application operating on the data processor carries out the following tasks:
	\begin{itemize}
		\item Generation of the \ac{jrc} waveform in digital, and forwarding them to the \ac{dfrc} Tx board for transmission.
		\item Processing the received radar echoes, implementing the scheme detailed in Subsection \ref{subsec:Radar}.
		\item Detection of the transmitted data symbol based on the received communications signal.
	\end{itemize}
	For experimental purposes, the application also provides the ability to embed a pre-defined target scheme into the received radar waveforms, allowing to evaluate the performance of the system with various configurable target profiles.
	
	The processing flow and the configuration of the setup parameters are controllable using a dedicated GUI, as detailed in the following subsection.

	\vspace{-0.2cm}
	\subsection{GUI: Configuration, Control and Display}
	\label{subsec:GUI}
\begin{figure*}
		\begin{minipage}[b]{1.0\linewidth}
			\centering
			\centerline{\includegraphics[width=0.8\columnwidth]{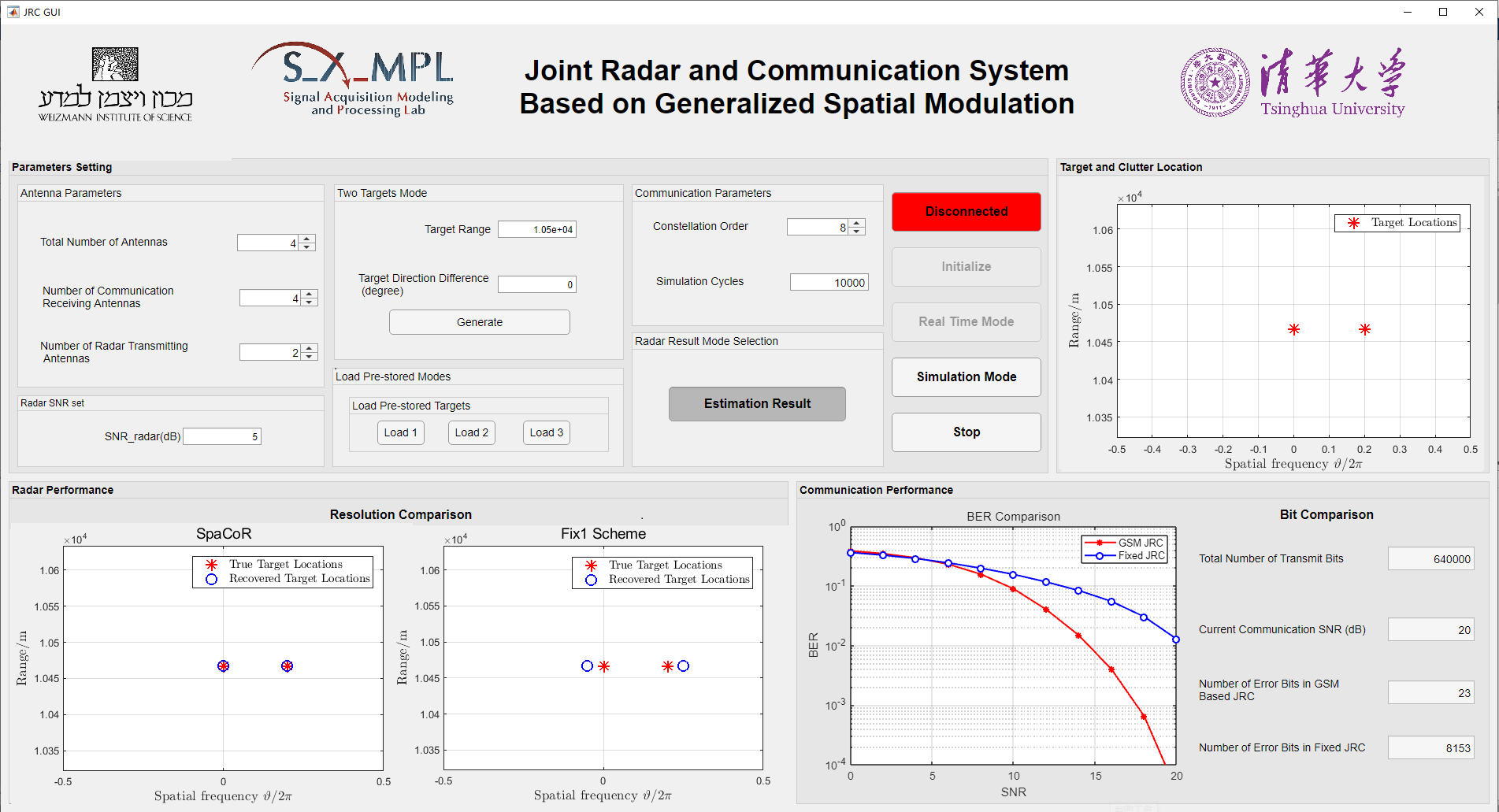}}
		\end{minipage}
		\vspace{-0.4cm}
		\caption{Graphical User Interface of the prototype}
		\vspace{-0.4cm}
		\label{fig:GUI}
		
	\end{figure*}

	\vspace{-0.1cm}
	A \ac{gui} is utilized to configure the prototype parameters, control the experiment process, and display the results. A screenshot of the GUI is shown in Fig. \ref{fig:GUI}.  
	
	\subsubsection{Parameter configuration} In order to simulate the \ac{dfrc} system using the hardware prototype, one must first select the system configuration. 
	%
	The configurable properties of the system include radar parameters, communication parameters, and \ac{dfrc} platform parameters. For the radar subsystem, the \ac{gui} allows to set the
	\ac{snr} of the radar echoes, i.e., the amount of noise added to the received waveforms in software, as well as the selection of the simulated target scenario mode. For the communications subsystem, one can specify the constellation order and the number of \ac{gsm} symbols used. For the platform parameters, the \ac{gui} allows configuring the number of elements used in the antenna array, i.e., $M$, as well as how many antenna elements are assigned for radar or communications, i.e., $M_{T}^{r}$ and $M_{T}^{c}$.
	
	\subsubsection{Controller}
	Once the parameters are configured, an experiment can be launched. The \ac{gui} allows the user to launch an experiment in two stages, by first initializing the hardware components to use the specified parameters, after which the transmission and reception can begin. Once the experiment is on-going, its results are updated in real-time, and is carried out until either all waveforms have been transmitted, or, alternatively, it is terminated by the user. 
	
	\subsubsection{Displayer}
	The experiment results  are visually presented by the \ac{gui} using three figures which are updated in real-time, as well as an additional static figure  displaying the  locations of the simulated targets. For the evaluation of radar performance, the \ac{gui} compares \ReviseDing{{\ac{spacora}}} with \emph{Fix1} by dedicating a figure to each scheme. These figures can compare either the beam pattern, or the target recovery resolution. For communication evaluation, the \ac{ber} curves of both methods are compared when transmitting at the same bit rate, i.e., the same number of bits per time slot.

	\vspace{-0.2cm}
	\section{Numerical Evaluations}
	\label{sec:Sims}
	\vspace{-0.1cm}
	In this section we evaluate  \ReviseDing{ \ac{spacora}} and compare it to \ac{dfrc} methods with fixed antenna allocation in hardware experiments and simulations. 
	The numerical evaluation of the radar and communications subsystems are detailed in Subsections \ref{subsec:RadarSims}-\ref{subsec:CommSims}, respectively. In particular, the radar performance in detecting multiple targets detailed here is based on the hardware prototype, while the remaining evaluations are carried out in simulations.  
	Table \ref{tab:ParameterValue} lists some of the parameters used in our prototype-aided experiments. While the prototype allows using $8$ antennas, in our experiments we use $M=4$ elements. 
	\begin{table}  
		\caption{Experiment settings.}
		\begin{center}  
			\begin{tabular}{||c|c|| c | c||}  
				\hline 
				\hline 
				Parameter & Value & Parameter & Value\\
				\hline 
				$M$ &  4 & $T_c$ & 2.5 $\mu$s\\
				$M_{T}^{r}$ & 2 & 	$T_r$ & 30 $\mu$s\\
				\hline 
				\hline		
			\end{tabular}  
		\end{center} 
		\label{tab:ParameterValue} 
	\end{table}   
	
	\vspace{-0.2cm}
	\subsection{Radar Subsystem Evaluation}
	\label{subsec:RadarSims}
	\vspace{-0.1cm}
	The analysis  in Section \ref{sec:RadarPerformance} indicates that 
	{\ac{spacora}} outperforms fixed antenna allocation schemes in several aspects: It has finer angular resolution compared with \emph{Fix1} and has lower sidelobes compared with \emph{Fix2}. In the following we demonstrate that these theoretical conclusions are also evident in our experiments, We first compare the angular resolution of \ReviseDing{{\ac{spacora}}} to \emph{Fix1} by comparing their ability to recover the locations of multiple adjacent targets. Then, the angle of a radar target is estimated in the presence of interference caused by clutters, allowing us to evaluate the sidelobe levels of the different \ac{dfrc} methods.
	
	\ReviseRadar{In the sequel, the bandwidth of radar waveform is set to $B_{r} = 50$ MHz.} \ReviseRadar{The locations of radar targets are recovered by using \ac{omp}\cite{eldar2012compressed} to solve  \eqref{eqn:RadarEquCS}. The delay interval and the spatial frequency interval are set to $\delta \tau = \frac{1}{5B_{r}}$ and $\delta \vartheta = \frac{2\pi}{5M}$, respectively.}

	\subsubsection{Radar Angular Resolution}
	{The angular resolution determines the smallest angular distance required to distinguish two adjacent targets located in the same range cell. It is   computed as half the width of the first two null points around the mainlobe of the beam pattern in the angular dimension.  
		From the analysis in Subsection \ref{subsec:comparison}, the angular resolution of \ReviseDing{{\ac{spacora}}} 
		is 
		$\frac{2\pi}{M}$, which equals 
		the angular resolution when using the full \ac{ula} with $M$ transmit antennas solely for radar. The resolution of \emph{Fix1} is 
		$\frac{2\pi}{M_{T}^{r}}$, which is larger than that of \ReviseDing{{\ac{spacora}}}.}
	%
	To demonstrate that this advantage of \ReviseDing{{\ac{spacora}}} over \emph{Fix1} is translated to improved target recover, we consider a scenario with two targets located in the same range cell but with different angular directions, where the reflective parameters $\{\alpha_l\}$ are all set to one, and the radar \ac{snr}, defined as  ${\rm{SNR}}^{\left(r\right)} := \frac{\left(M_{T}^{r}\right)^2 P_{t}}{\sigma_{r}^{2}}$, is set to $0$ dB. The direction of transmit beam is set to  $\theta_{T} = 0$.  Specifically, in the considered scenario the angular difference between the two adjacent targets is larger than the angular resolution of \ReviseDing{{\ac{spacora}}} while smaller than the angular resolution of \emph{Fix1}.   
	\begin{figure}		
		\begin{minipage}[b]{\linewidth}
			\centering
			\centerline{\includegraphics[width=\subfigWidth]{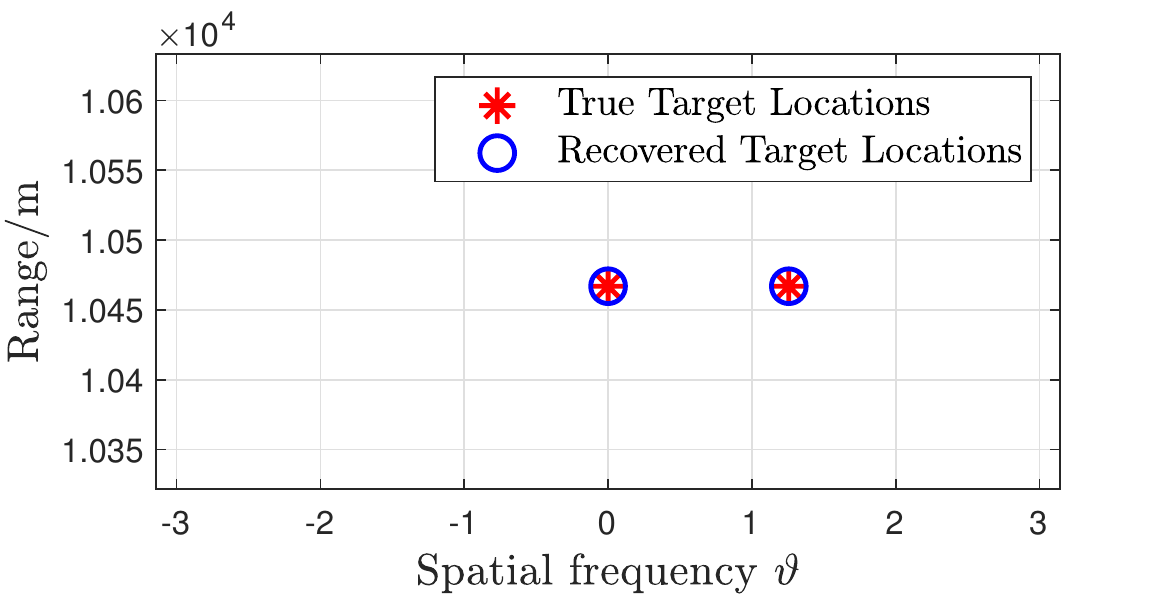}}
			\centerline{\small (a) \ReviseDing{\ac{spacora}}}\smallskip
		\end{minipage}
		
		\begin{minipage}[b]{\linewidth}
			\centering
			\centerline{\includegraphics[width=\subfigWidth]{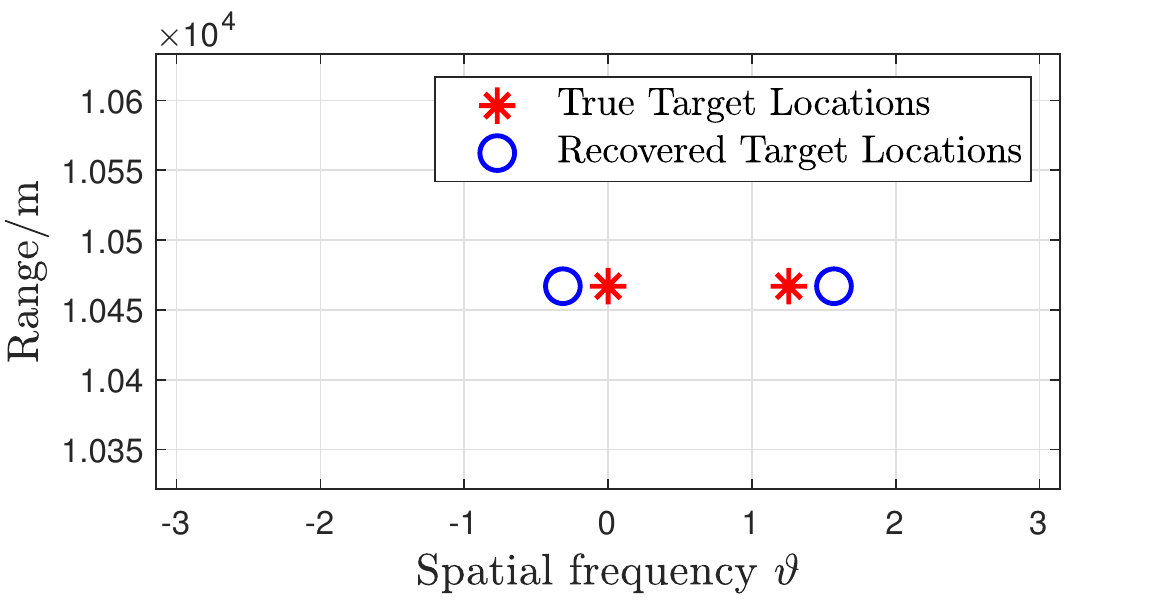}}
			\centerline{\small(b) Fix1}\smallskip
		\end{minipage}	
		\vspace{-0.6cm}
		\caption{\ReviseRadar{Recovery results for two adjacent targets.}}
		\label{fig:RadarResoSimu1}
	\end{figure}
	
	\begin{figure}		
		\begin{minipage}[b]{\linewidth}
			\centering
			\centerline{\includegraphics[width=\subfigWidth]{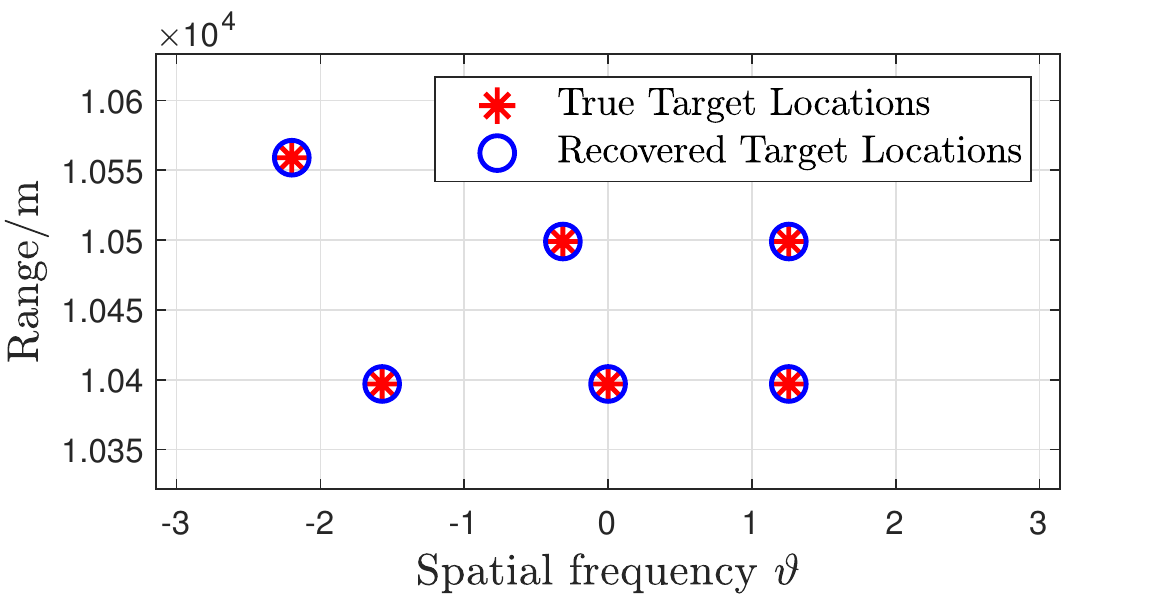}}
			\centerline{\small(a) \ReviseDing{{ \ac{spacora} }}}\smallskip
		\end{minipage}
		
		\begin{minipage}[b]{\linewidth}
			\centering
			\centerline{\includegraphics[width=\subfigWidth]{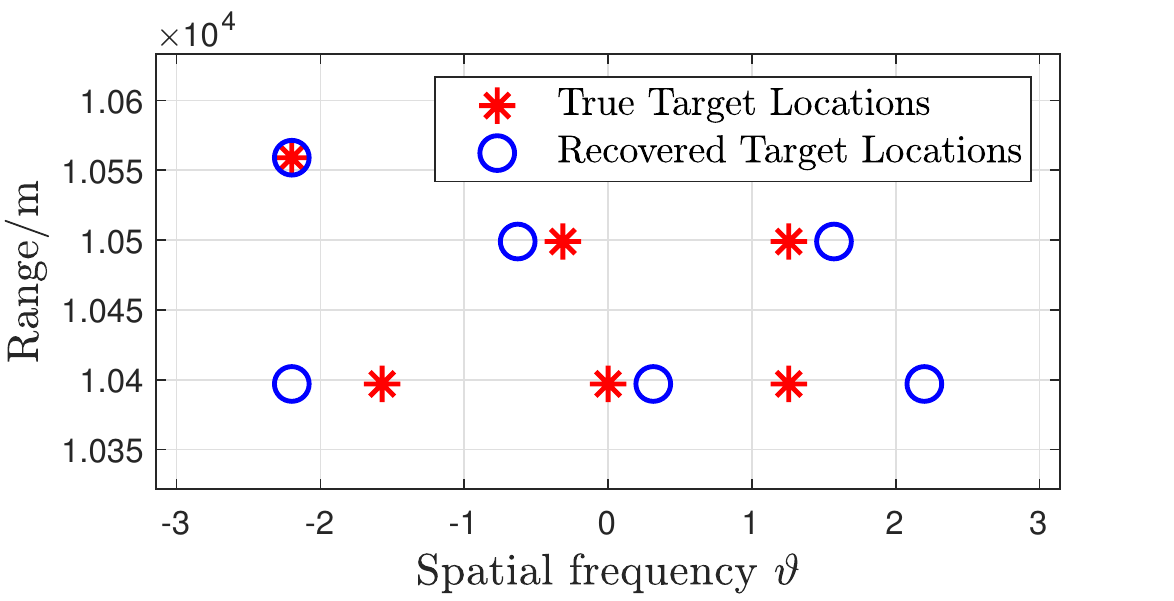}}
			\centerline{\small(b) Fix1}\smallskip
		\end{minipage}	
		\vspace{-0.6cm}	
		\caption{\ReviseRadar{Recovery results for six targets.}}
		\label{fig:RadarResoSimu2}
	\end{figure}
	The recovery results of \ReviseDing{{\ac{spacora}}} and \emph{Fix1} are shown in Fig.\ref{fig:RadarResoSimu1}(a)-\ref{fig:RadarResoSimu1}(b), respectively, along with the true locations of the targets.  \ReviseRadar{From the recovery results, we observe that  {\ac{spacora}} is capable of accurately recovering the target locations, while \emph{Fix1} fails to do so.} This is because that {\ac{spacora}} has improved resolution than that of \emph{Fix1}. 
	The recovery results of a more complex scenario with six targets are depicted in Fig.~\ref{fig:RadarResoSimu2}, which further demonstrates the improved ability of \ReviseDing{{\ac{spacora}}} in identifying multiple adjacent targets.
	

	\subsubsection{Sidelobe Level}
	Radar target detection performance is degraded in the presence of clutters, where the magnitude of this degradation is related to the sidelobe level of the transmit beam pattern. A higher sidelobe level radiates more energy in the direction of the clutters,  resulting in increased interference which in turn degrades detection performance. In Section \ref{sec:RadarPerformance}, we analyzed the sidelobe levels of \ReviseDing{ \ac{spacora}} and the fixed allocation schemes through the variance of transmit beam pattern. In this simulation, we demonstrate that the reduced sidelobe levels of \ac{spacora} translate into improved target detection accuracy. To that aim, we consider a scenario in which one radar target is located in the mainlobe of the radar transmit beam pattern, and evaluate radar performance in the presence of clutter. \ReviseRadar{Two clutters are randomly generated outside the mainlobe of the transmit beam pattern in the same range cell with the radar target. The amplitudes of the clutter reflective factors are randomized following a Rayleigh distribution as in \cite[Ch. 2.2]{Richards2005fundamentals}.}
	
\begin{figure}			
		\centerline{\includegraphics[width=0.8\columnwidth]{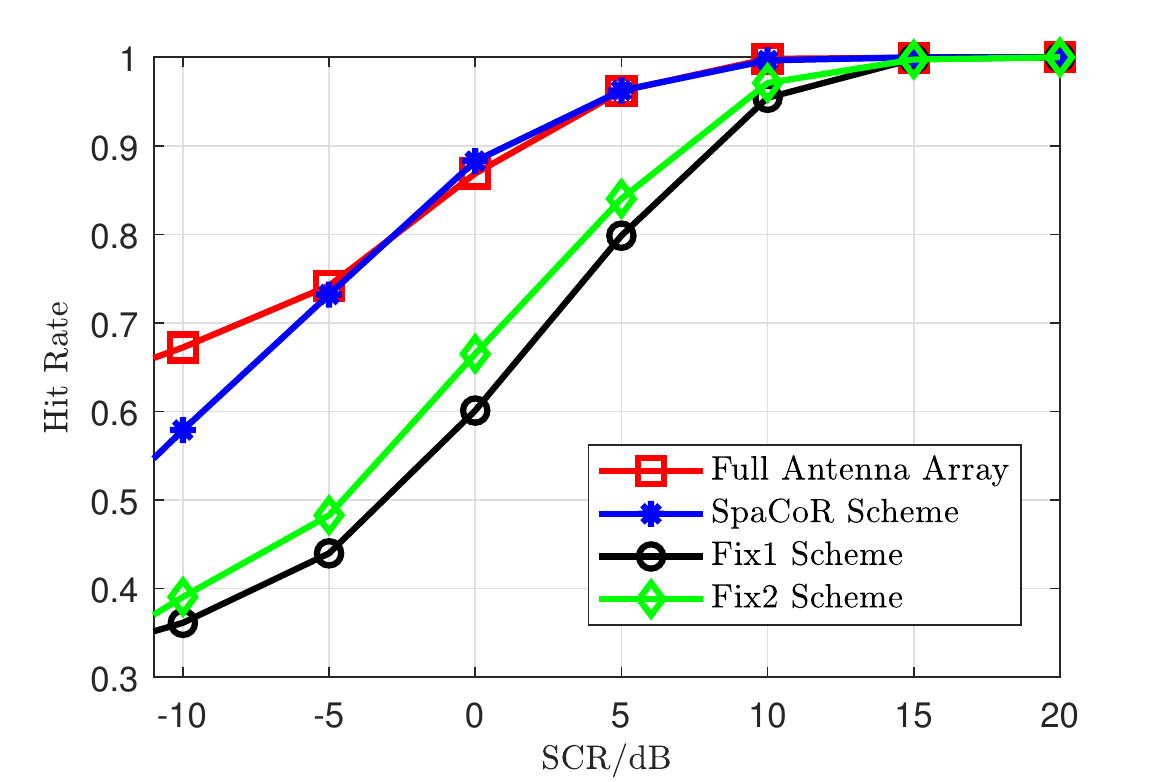}}
		\vspace{-0.2cm}
		\caption{\ReviseRadar{Hit Rate of angle estimate in different antenna allocation schemes.}}
		\label{fig:AntiClutter}
	\end{figure}

	\ReviseRadar{Here, we use hit rate as the performance criterion. A ”hit”
		is defined if the angle parameter of the target is  successfully recovered. The hit rates of angle estimates are calculated over $4000$ Monte Carlo trials versus the 
		\ac{scr}, defined as the ratio between the square of target reflective factor and the square of the expected clutter reflective factor. The results are depicted  in Fig. \ref{fig:AntiClutter}, where the hit rate curves of the full antenna array, {{\ac{spacora}}}, \emph{Fix1} and \emph{Fix2} are shown. Observing these hit rate curves, 
		we find that the hit rate of  {{\ac{spacora}}} approaches that of using the full array, and outperforms the fixed allocation methods, while \emph{Fix1} achieves the lowest hit rate for all considered \ac{scr} values.} These results are in line with the theory analysis detailed in Section \ref{sec:RadarPerformance}: Due to the shrinkage of antenna aperture, the mainlobe of \emph{Fix1} is wider than the mainlobes of {{\ac{spacora}}} and \emph{Fix2}. Hence, the interference introduced by the clutters are strongest, which notably degrades the radar performance. {{\ac{spacora}}} outperforms \emph{Fix2} as the variance of transmit beam pattern for {{\ac{spacora}}} is lower than that of \emph{Fix2}, and thus its sidelobe levels are lower and it is less sensitive to interference caused by clutters. 
	
	\vspace{-0.2cm}
	\subsection{Communications Subsystem Evaluation}
	\label{subsec:CommSims}
	\vspace{-0.1cm}
	The communications subsystem of \ReviseDing{{\ac{spacora}}}  is based on \ac{gsm} signaling. For comparison, the \ac{dfrc} systems with fixed antenna allocation do not encode bits in the selection of the antennas, and thus convey their information only via conventional \ac{smx}.  To compare the communication capabilities of the considered methods, we compare their uncoded \ac{ber} performance, 
	To that aim, a total of $10^5$ \ac{jrc} waveforms are transmitted and decoded by the receiver, To guarantee fair comparison,   we set the data rates of the considered methods to be identical. This is achieved by using constellations of different orders. In particular, we compare GSM-QPSK, which conveys two spatial bit in the selection of the two antennas from an array of $M=4$ elements, and four constellation bits, in the form of two QPSK symbols, per \ac{gsm} symbol, with SMX-8PSK. Similarly, we compare the \ac{ber} achieved when using  GSM-8PSK  to that of  SMX-16PSK, both  of  which  transmit  8 bits in each time slot.
	
\begin{figure} 		
		\centerline{\includegraphics[width=0.8\columnwidth]{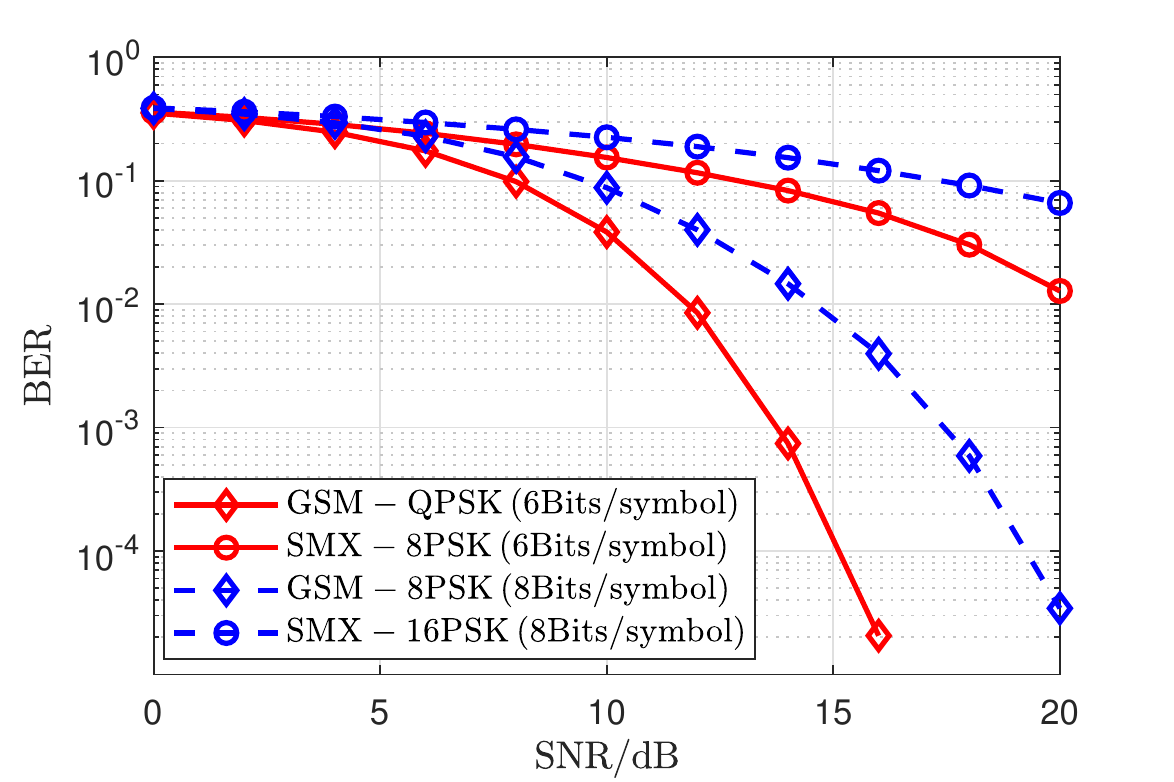}}
		\vspace{-0.4cm}
		\caption{BER comparisons of \ac{gsm} and \ac{smx}.}
		\label{fig:BER}
	\end{figure}

	The \ac{ber} curves of the \ac{gsm}-based { \ac{spacora}} compared to \ac{dfrc} systems with fixed antenna allocation utilizing \ac{smx} for communications are depicted in Fig. \ref{fig:BER}. 
	We observe in  Fig. \ref{fig:BER} that for the same data rates,  \ac{gsm} achieves improved \ac{ber} performance compared to \ac{smx}, and that  its \ac{ber} curve decreases faster than \ac{smx} with \ac{snr}. This gain follows from the fact that \ac{gsm} utilizes less dense constellations compared to \ac{smx}, as it conveys additional bits in the selection of the antenna indices. Nonetheless, this performance gain comes at the cost of increased decoding complexity at the receiver, which is a common challenge associated with \ac{im} schemes.

\begin{figure}		
		\centering
		\centerline{\includegraphics[width=0.8\columnwidth]{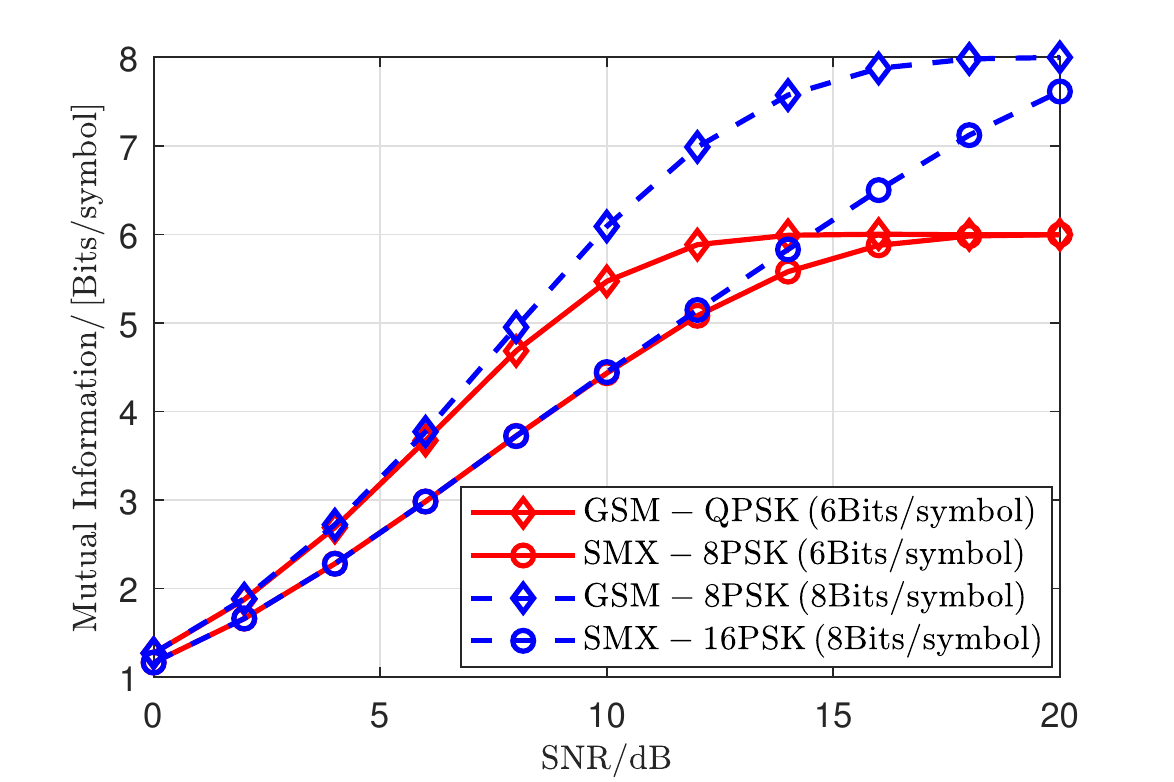}}
		\vspace{-0.4cm}
		\caption{Mutual information comparisons of \ac{gsm} and \ac{smx}.}
		\label{fig:MI}
	\end{figure}

	The communication performance gains of \ac{gsm}, observed in our evaluation of its uncoded \ac{ber} performance, are also evident when evaluating its \ac{mi} between the transmitted signal and the channel output, which represents its achievable rate.  
	%
	To demonstrate this gain, we numerically compare the \ac{mi} of \ac{gsm} with that of \ac{smx} in Fig. \ref{fig:MI}. Our  evaluation of the \ac{mi} values are based on the derivation of the \ac{mi} of \ac{gsm} given in \cite{Younis2018Information}.
	Observing Fig. \ref{fig:MI}, we note that, as expected, the \ac{mi} does not exceed the number of bits encapsulated in each symbol. Consequently, the maximal \ac{mi} of GSM-QPSK and SMX-8PSK equal to 6 bits per symbol, while the maximal \ac{mi} of GSM-8PSK and SMX-16PSK equals 8 bits per symbol. However, these data rates can only be achieved reliably at high \ac{snr} values. In lower \acp{snr}, \ac{gsm} achieves improved \ac{mi} over \ac{smx}, indicating that it is capable of reliably conveying larger volumes of data. These results, combined with the radar performance evaluated in Subsection \ref{subsec:RadarSims}, demonstrate that the usage of \ac{gsm} in \ac{dfrc} systems contributes to both radar, in its introduction of radar agility which contributes to the angular resolution, as well as the communications subsystem, allowing it to achieve improved performance in terms of both \ac{ber} as well as achievable rate.
	
	\vspace{-0.2cm}
	\section{Conclusions}
	\label{sec:Conclusions}
	In this paper, we proposed \ReviseDing{{\ac{spacora}}}, which is a \ac{dfrc} system based on \ac{gsm}.  \ReviseDing{{\ac{spacora}}} conveys additional bits  by the combinations of transmit antenna elements, and the antenna allocation patterns change between symbols in a random fashion introducing spatial agility. The signal models and processing algorithms  were presented. In order to evaluate the radar performance, we characterized the transmit beam pattern and analyzed its stochastic performance, showing that the beam pattern of the proposed system approaches that   of using the full antenna array solely for radar. To demonstrate the feasibility of the approach, we built a dedicated hardware prototype realizing this \ac{dfrc} system using over-the-air signaling.  Hardware experiments and simulations demonstrated the gains of the proposed method over \ac{dfrc} systems using fixed antenna allocations in terms of both radar resolution and sidelobe level, as well as communication  \ac{ber} and achievable rate. Our results  and the presented hardware prototype narrow the gap between the theoretical concepts of \ac{im}-based \ac{dfrc} systems and their implementation in practice.

	\numberwithin{proposition}{subsection} 
	\numberwithin{lemma}{subsection} 
	\numberwithin{corollary}{subsection} 
	\numberwithin{remark}{subsection} 
	\numberwithin{equation}{subsection} 
	
	\vspace{-0.2cm}
	\begin{appendix}
		
		\vspace{-0.2cm}
		\subsection{Proof of Theorem \ref{thm:ExpBeamPattern}}
		\label{app:Proof1}
		\vspace{-0.1cm}
		For a given time slot $k$, the indices of the radar transmitting antennas, denoted by  ${\mathbf{M}}_k^{\rm GSM}$, are randomized from the radar antenna combination set. The random vector ${\mathbf{M}}_k^{\rm GSM}$ thus obeys a discrete uniform distribution over this set, i.e., $\Pr\left({\mathbf{M}}_k^{\rm GSM}\right)=1/P_{c}$, where $P_{c} := \binom{M}{M_{T}^{r}}$ is the total number of possible antenna index combinations.
		The expected  transmit delay-direction beam pattern can be calculated as follows: 
		\ReviseDing{
			\begin{flalign}
			&\mathcal{E}\!\left\{\chi_{T}^{\rm GSM}\left(\tau_{d}, f_{\theta}\right)\right\} = \sum_{\dtidx = 0}^{\dtmax-1}\sum_{k=0}^{K-1} \mathcal{E}\!\left\{\rho_T\left(k, f_{\theta}\right)\right\} \notag \\
			&\quad \quad \times g\!\left(\!\frac{\dtidx T_{s}\! - \!kT_c\! - \tau_{d} - \tilde{\tau}}{T_c}\!\right)\!\!h\!\left[\dtidx , \tau_{d} + \tilde{{\tau}}\right] \!h^{\ast}\left[\dtidx , \tilde{\tau}\right].
			\label{eqn:proofexpectation1}
			\end{flalign}
			The} expected value of the transmit gain in \eqref{eqn:proofexpectation1} is \ReviseDing{
			\begin{align}
			&\mathcal{E}\left\{\rho_T\left(k, f_{\theta}\right)\right\} = \mathcal{E}\left\{\sum_{l=0}^{M_{T}^{r}-1}e^{j m_{k,l}f_{\theta}}\right\} \notag \\
			& \stackrel{(a)}{=} \frac{1}{P_{c}} \sum_{i=0}^{P_{c}-1}\sum_{l=0}^{M_{T}^{r}-1}e^{j  m_{l}^{(i)} f_{\theta}} 
			\stackrel{(b)}{=} \frac{1}{P_{c}} \!\cdot\! \frac{P_{c} \!\cdot\! M_{T}^{r}}{M} \!\sum_{m=0}^{M-1} \! e^{jm f_{\theta}} \notag \\
			& = e^{-j\frac{f_{\theta}}{2}}\cdot \frac{M_{T}^{r}}{M}\cdot\frac{\sin\left(Mf_{\theta}/2\right)}{\sin\left({f_{\theta}/2}\right)},
			\label{eqn:ProofExpTranGain}
			\end{align}
			where} $(a)$ follows since $\mathbf{M}_k^{\rm GSM}$ is uniformly distributed, and $(b)$ holds as there are $P_{c}\cdot M_{T}^{r}$ items in the summation, where each index in  $\left\{0, 1, \ldots, M-1\right\}$ occurs $P_{c}M_{T}^{r}/M$ times. As $\mathcal{E}\left\{\rho_T\left(k, f_{\theta}\right)\right\}$ does not depend on the index $k$, we substitute $\mathcal{E}\left\{\rho_T\left(k, f_{\theta}\right)\right\}$ by $\mathcal{E}\left\{\rho_T\left(\cdot, f_{\theta}\right)\right\}$, and \eqref{eqn:proofexpectation1} is rewritten as
		\ReviseDing{
			\begin{align}
			&\mathcal{E}\!\left\{\chi_{T}^{\rm GSM}\left(\tau_{d}, f_{\theta}\right)\right\} =\mathcal{E}\!\left\{\rho_T\left(\cdot, f_{\theta}\right)\right\} \notag \\
			& \times \sum_{\dtidx=0}^{\dtmax-1} \sum_{k=0}^{K-1}  g\!\left(\!\frac{\dtidx T_{s}\! - \!kT_c\! - \tau_{d} - \tilde{\tau}}{T_c}\!\right)h\!\left[\dtidx , \tau_{d} + \tilde{{\tau}}\right] \!h^{\ast}\left[\dtidx , \tilde{{\tau}}\right] \notag  \\
			& = \mathcal{E}\left\{\rho_T\left(\cdot, f_{\theta}\right)\right\}\cdot \sum_{\dtidx = 0}^{\dtmax-1} h\!\left[\dtidx , \tau_{d} + \tilde{{\tau}}\right] h^{\ast}\left[\dtidx , \tilde{{\tau}}\right] ,
			\label{eqn:proofexpectation2}
			\end{align}
			which follows from $\sum_{k=0}^{K-1}  g\big(\!\frac{\dtidx T_{s} - kT_c - \tau_{d} - \tilde{\tau}}{T_c}\!\big)  = g\big(\frac{\dtidx T_{s} - \tau_{d} - \tilde{\tau}}{T_{r}}\big)$, and since $h\left(t\right)$ is a pulse with width $T_{r}$, $g\big(\frac{\dtidx T_{s} - \tau_{d} - \tilde{{\tau}}}{T_{r}}\big)h\left[\dtidx , \tau_{d} + \tilde{{\tau}}\right] = h\left[\dtidx , \tau_{d} + \tilde{{\tau}}\right]$. }  
		Substituting \eqref{eqn:ProofExpTranGain} into \eqref{eqn:proofexpectation2} and taking its absolute values proves \eqref{eqn:ExpBeamPattern}.\qed

		
		\vspace{-0.2cm}
		
		\subsection{Proof of Proposition \ref{pro:Variance}}
		\label{app:Proof2}
		\vspace{-0.1cm}
		The variance of the transmit beam pattern is
		\begin{align}
		\mathcal{V} 
		& =\mathcal{E}\left\{\left|\chi_{T}^{\rm GSM}\left(\tau_{d}, f_{\theta}\right)\right|^2\right\} \!-\! \left|\mathcal{E}\left\{\chi_{T}^{\rm GSM}\left(\tau_{d}, f_{\theta}\right)\right\}\right|^2.
		\label{eqn:ProofVariance1}
		\end{align}
		The second term in \eqref{eqn:ProofVariance1} is given in  \eqref{eqn:ExpBeamPattern}. 
		By defining 
		\ReviseDing{\begin{align*}
			\!\!\!\eta\left(k, \tau_{d}\right)  \!= \!\sum_{\dtidx=0}^{\dtmax-1} g\!\left(\!\frac{\dtidx T_{s} \!-\! kT_c \!-\! \tau_{d} - \tilde{{\tau}}}{T_c}\right)\!h\left[\dtidx , \tau_{d} \!+\! \tilde{{\tau}}\right] h^{\ast}\left[\dtidx , \tilde{{\tau}}\right], \!\!
			\end{align*}	
			it} can be shown that  $\mathcal{E}\Big\{\big|\chi_{T}^{\mathrm{GSM}}\big(\tau_{d}, f_{\theta}\big)\big|^2\Big\}$  equals $\mathcal{E}\big\{\sum_{k=0}^{K-1}\sum_{k'=0}^{K-1}\big[\eta \big(k,\tau_{d}\big)\rho_T\big(k,f_{\theta}\big) \eta^{\ast}\big(k',\tau_{d}\big)\rho_T^{\ast}\big(k', f_{\theta}\big)\big]\big\}$, and can thus be written as
		\begin{align}
		&\mathcal{E}\left\{\left|\chi_{T}^{\mathrm{GSM}}\left(\tau_{d}, f_{\theta}\right)\right|^2\right\} = \sum_{k=0}^{K-1}\!\left|\eta\left(k,\tau_{d}\right)\right|^2\mathcal{E}\!\left\{\!\left|\rho_T\!\left(k,f_{\theta}\right)\right|^2\!\right\}\! + \notag \\
		& \!\sum_{k=0}^{K-1}\!\sum_{  k' \ne {k} } \eta\left(k,\tau_{d}\right)\eta^{\ast}\!\left(k',\tau_{d}\right)\!\mathcal{E}\!\left\{{\rho_{T}\!\left(k,f_{\theta}\right)\!\rho_{T}^{\ast}\left(k',f_{\theta}\right)}\right\}.
		\label{eqn:Proof2rdMoment}
		\end{align}
		To compute \eqref{eqn:Proof2rdMoment}, we note that by \eqref{eqn:RhoT} and the fact that $\mathbf{M}_{k}^{\rm GSM}$ obeys a uniform distribution, it holds that
		\begin{align}
		&\mathcal{E}\left\{\left|\rho_{T}\left(k,f_{\theta}\right)\right|^2\right\} 
		=\frac{1}{P_{c}} \sum_{i = 0}^{P_{c}-1}\sum_{l=0}^{M_{T}^{r}-1}\sum_{l' = 0}^{M_{T}^{r}-1}e^{j\left(m_{l}^{(i)} - m_{l'}^{(i)}\right)f_{\theta}} \notag \\
		& \stackrel{(a)}{=} \frac{1}{P_{c}} \!\!\left\{\!\frac{P_{c}M_{T}^{r}\left(M_{T}^{r} \!-\! 1\right)}{M\left(M-1\right)}\!\left|\!\sum_{m=0}^{M-1}\!e^{jm f_{\theta}}\!\right|^{2} \!\!\!+\! \frac{P_{c}M_{T}^{r}\left(M \!-\! M_{T}^{r}\right)}{M-1}\!\right\} \notag \\
		& = \! \frac{M_{T}^{r} \! \left(M_{T}^{r} \! - \! 1\right) \!}{M \!\left(M \! - \!1\right)} \! \left|\frac{\sin\left(Mf_{\theta}/2\right)}{\sin\left(f_{\theta}/2\right)}\right|^2   +  \frac{M_{T}^{r}\left(M \! - \! M_{T}^{r}\right)}{M \! - \!1},
		\label{eqn:ProofVarianceterm2}
		\end{align}
		where 
		$(a)$ holds as the summation can be decomposed into constant terms, which add up to $\frac{P_{c}M_{T}^{r}\left(M - M_{T}^{r}\right)}{M-1}$, and to the term $\left|\sum_{m=0}^{M-1}e^{j m f_{\theta}}\right|^{2}$, which repeats $\frac{P_{c}M_{T}^{r}\left(M_{T}^{r} - 1\right)}{M\left(M-1\right)}$ times. Additionally, for $k \neq k'$ the random variables $\rho_{T}\left(k, f_{\theta}\right)$ and $\rho_{T}^{\ast}\left(k', f_{\theta}\right)$ are independent, and thus  
		\begin{align}
		\!\!&\mathcal{E}\left\{\rho_{T}\left(k, f_{\theta}\right)\rho_{T}^{\ast}\left(k', f_{\theta}\right)\right\} 
		\!= \!
		\left(\frac{M_{T}^{r}}{M}\right)^2\!\left|\frac{\sin\left(Mf_{\theta}/2\right)}{\sin\left(f_{\theta}/2\right)}\right|^2.
		\label{eqn:ProofVarianceterm3}
		\end{align}
		Furthermore, it holds that \ReviseDing{
			$\sum_{k=0}^{K-1}\sum_{ k' \!\ne\! {k} }\eta\big(k,\tau_{d}\big)\eta^{\ast}\big(k'\!,\tau_{d}\big) 
			\!=\! \big|\sum_{\dtidx=0}^{\dtmax-1}h\left[\dtidx , \tau_{d} + \tilde{\tau}\right] h^{\ast}\left[\dtidx , \tilde{{\tau}}\right] \big|^2 \! - \!\sum_{k = 0}^{K-1}\!\left|\eta\left(k, \tau_{d}\right)\right|^2$.} 
		Substituting this as well  
		into \eqref{eqn:ProofVariance1},  we obtain 
		\begin{flalign}
		\!&\mathcal{V}\left\{\chi_{T}^{\rm GSM}\left(\tau_{d}, f_{\theta}\right)\right\} =  \sum_{k=1}^{K}\left|\eta\left(k,\tau_{d}\right)\right|^2 \times \notag \\ &\left\{\!\!\frac{M_{T}^{r}\left(M_{T}^{r} \!-\! M\right)}{M^{2}\left(M-1\right)} \!\!\left|\frac{\sin\left(Mf_{\theta}/2\right)}{\sin\left(f_{\theta}/2\right)}\right|^2 \!\!\!+ \!\frac{M_{T}^{r}\left(M \!-\! M_{T}^{r}\right)}{M - 1} \!\!\right\}\!. \!\!
		\label{eqn:ProofVariance2}
		\end{flalign}
		When we specialize in \eqref{eqn:Chirp1} for chirp waveforms, it holds that
		\ReviseDing{$\sum_{k=0}^{K-1}\left|\eta\left(k,\tau_{d}\right)\right|^2\ \approx \frac{\dtrad^2}{K}\mathrm{sinc}^{2}\left(\frac{B_r\tau_{d}}{K}\right)$.} Utilizing this together with  \eqref{eqn:ExpBeamPattern_chirp}  proves  \eqref{eqn:VarianceBeamPattern}.
		\qed 
		
	\end{appendix}

	\bibliographystyle{IEEEtran}
	\bibliography{IEEEabrv,references}

\end{document}